\documentstyle[11pt,aaspp4]{article}
\def\msun{M_{\odot}}
\begin{document}
\slugcomment{Accepted for publication in the Astrophysical Journal}
\title{Thermal Equilibrium Curves and Turbulent Mixing
in Keplerian Accretion Disks}
\author{Pin-Gao Gu}
\affil{Dept. of Physics, University of Texas, Austin, TX 78712;
gu@physics.utexas.edu}
\author{Ethan T. Vishniac}
\affil{Dept. of Physics and Astronomy, Johns Hopkins University,
        Baltimore, MD 21218; ethan@pha.jhu.edu}
\and
\author{John K. Cannizzo}
\affil{Goddard Space Flight Center, NASA/GSFC/Laboratory for
        High Energy Astrophysics, Code 662, Greenbelt, MD 20771;
cannizzo@stars.gsfc.nasa.gov}

\begin{abstract}
We consider vertical heat transport in Keplerian accretion
disks, including the effects of radiation, convection, and 
turbulent mixing driven by the Balbus-Hawley instability, in 
astronomical systems ranging from dwarf novae (DNe), and 
soft X-ray transients (SXTs), to active galactic nuclei (AGN). 
In order to account for the interplay between convective and 
turbulent energy transport in a shearing environment, we 
propose a modified, anisotropic form of mixing-length theory,
which includes radiative and turbulent 
damping.  We also include turbulent heat transport, which acts
everywhere within disks, regardless of whether or not they
are stably stratified, and can move entropy in either 
direction.  We have generated a series of vertical structure 
models and thermal equilibrium curves using the scaling law for 
the viscosity parameter $\alpha$ suggested by the 
exponential decay of the X-ray luminosity in SXTs.  We have 
also included equilibrium curves for DNe using an $\alpha$ 
which is constant down to a small magnetic Reynolds number ($\sim 10^4$).
Our models indicate that weak convection is usually eliminated by
turbulent radial mixing, even when
mixing length estimates of convective vertical heat transport
are much larger than turbulent heat transport.
The substitution of turbulent heat transport for convection
is more important on the unstable branches of thermal equilibrium S-curves
when $\alpha$ is larger. The low temperature turnover
points $\Sigma_{max}$ on the equilibrium S-curves
are significantly reduced by turbulent mixing in DNe and
SXT disks. However, in AGN disks the standard mixing-length 
theory for convection is still a useful approximation when we
use the scaling law for $\alpha$, since these disks
are very thin at the relevant radii.  In accordance
with previous work, we find that constant $\alpha$ models give
almost vertical S-curves in the $\Sigma-T$ plane and consequently 
imply very slow, possibly oscillating, cooling waves.
\end{abstract}


\section{Introduction}

The disk-instability model in a thin accretion disk
is the most popular model for
dwarf-novae (DNe) outbursts and soft X-ray transients (SXTs)
(for recent reviews see \cite{can93} or \cite{osa96}).
In this model the initial fast rise of luminosity in
an outburst is due to the propagation of a heating front
at the local thermal speed, while the exponential decay follows
from the subsequent slow propagation of a cooling
wave.  The alternation of heating and cooling in
these systems comes from the characteristic `S' shaped
curve in the midplane temperature versus column density
equilibrium curve (see Fig.~ \ref{fig1} below).  
At all points along this curve the heating due to the
local dissipation of orbital energy is balanced by radiative
losses and convective heat transport.  The unstable
middle branch of the equilibrium curve, where column density
falls with increasing midplane temperature, is produced
by the hydrogen ionization transition and its effect on
radiative opacities.   The local heating rate is usually
taken to follow some variant of the
$\alpha$ prescription (\cite{ss73}).
Vertical heat transport in accretion disks is
modeled following the usual treatment for radial
heat transport in stars, with radiation and convection
defining the thermal energy balance. 
Unlike main-sequence stars which have stable convective
zones and internal veloicty fields, convection in
accretion disks varies with time as they evolve along the 
thermal equilibrium curves.
Typically, convection has a very strong influence on the 
structure of the thermally and viscously unstable middle
branch of the thermal equilibrium curve since the optically
depth of a partially ionized disk is very large.

Convective mixing is not the only source of turbulent transport
in stars and disks.
Diffusive mixing induced by shear instabilities in rotating stars
has been examined by several researchers (\cite{zah92},
\cite{mae95}, \cite{mae96}) for its effects on
the luminosity,  and evolution of stars.  Similarly, it is
appropriate to include shear-induced turbulent mixing
in the study of accretion disk structure, especially since
the luminosity of accretion disks is ultimately due to the dissipation
of orbital energy through some sort of shear instability.  Turbulent
mixing is an unavoidable part of this process.  Furthermore,
a particular local instability  has been identified (Velikhov 1959,
Chandrasekhar 1960, Balbus and
Hawley 1991, Hawley and Balbus 1992) which is uniquely
suited to playing this role, and consequently gives us a theoretical
basis for evaluating the efficiency of turbulent mixing.
In accretion disks the Velikhov-Chandrasekhar instability is
commonly referred to as the Balbus-Hawley instability. Here we
abbreviate this as the BH instability.
In contrast to convective transport, which only affects
the convectively unstable regions and moves heat
down temperature gradients, turbulent transport
driven by the BH instability is ubiquitous within magnetized
accretion disks, and will always act to reduce entropy
gradients.  The effect of turbulent diffusion
due to some sort of shearing instability, not necessarily 
the BH instability, has been studied in the context
of radial transport in thin (important for transition
fronts, e.g. see \cite{mhs99})
and advective (\cite{h96}) disks, and 
in the context of vertical transport in
T Tauri disks (\cite{DAlessio}).

In a real disk the BH instability should generate
turbulence which is only moderately anisotropic (\cite{vd92}),
an expectation which is consistent with the available
numerical simulations (see Gammie 1997 and references contained
therein).  Consequently, we expect
that the dimensionless form of all the transport
coefficients should differ only by factors of order unity.
More precisely, we note that the numerical simulations
consistently suggest that the angular momentum transport
is dominated by magnetic tension, and that the
relevant magnetic field moment $<B_rB_\theta>$ is a few times
$\rho<v_z^2>$.  The vertical diffusion coefficient
is $<v_z^2>\tau$, where $\tau$ is the velocity correlation
time scale, and $\Omega\tau$ is a factor of a few, where
$\Omega$ is the local orbital frequency.  We conclude that
the vertical diffusion coefficient is comparable to
the radial diffusion coefficient. In other words,
\begin{equation}
D_z\sim \nu,
\label{eq:vert}
\end{equation}
where $\nu$ is the usual notation for the
turbulent kinematic viscosity.
Here we will assume that equation (\ref{eq:vert}) is an
exact equality, although in reality we expect that this is only a
rough approximation.  Consequently, if we assume that the
turbulent Prandtl number is unity, the vertical thermal
diffusion coefficient $\chi_T$ can be written as
\begin{equation}
\chi_T\sim \nu={2\over 3}{\alpha P\over \rho \Omega},
\label{conduct}
\end{equation}
where
$\Omega$ is the Keplerian angular velocity,
and $P$ and $\rho$
are the disk pressure and mass density.
The
turbulent mixing rate across the disk is
\begin{equation}
k_z^2 \chi_T \sim \alpha \Omega,
\end{equation}
where we have assumed a vertical wavelength $1/k_z$ comparable to
the vertical global pressure scale height $\lambda_P$.
Since this is also the thermal equilibration rate for the disk, 
this implies
that turbulent mixing can have a significant effect on disk structure.

The BH instability follows from the presence
of a magnetic field in the disk, which requires some sort
of magnetic dynamo.  The value of $\alpha$ is a fraction of
$(V_A/c_s)^2$, the ratio of magnetic pressure to ambient pressure.
This ratio is determined
by the balance between turbulent dissipation and flux generation
via the (uncertain) dynamo process.  It has long been known that
the observed luminosities and timescales for DNe
require a larger hot state viscosity,
$\alpha_{hot}$, than in quiescence, $\alpha_{cold}$
(\cite{s84}, \cite{ls91}).  This could indicate
a dependence of $\alpha$ on temperature, or more probably on the
disk geometry.  On the other hand, it could simply reflect the
fact that cold disks have a large neutral fraction and a high
resistivity.  The latter consideration has led Gammie and Menou 
(1998) to propose that disks in quiescence are cold enough
to shut down all dynamo activity.  Evidence in favor of
the former idea comes
from the observed exponential decay of soft X-ray
luminosity for soft X-ray transients which seems to require
a scaling law for $\alpha$ of the form (\cite{ccl95}, 
\cite{vw96},\cite{vis97})
\begin{equation}
\alpha \approx \alpha_0\left( {{c_s}\over r\Omega} \right)^{1.5},
\label{eq:ascale}
\end{equation}
where $\alpha_0=50$,
${c_s}$ is the isothermal sound speed at
the mid-plane, and $r$ is the radial position of the annulus.
The value of $\alpha_0$ is proportional to the assumed masses of the
central black holes used to calibrate this relation.  Cannizzo
et al. assumed a black hole mass of $10 M_{\sun}$.
This scaling law has been criticized (see e.g. \cite{kr98}
and references therein) on the grounds that irradiation should
substantially modify the structure and stability of disks
around black holes.  Certainly a naive calculation of the 
optical emission from the disks of SXTs during the decline
from outburst fails to match the observed behavior (\cite{csl97}).  
However, a recent study (\cite{dlhc99})
has shown that it is much more difficult for self-consistent
models of irradiation to change the basic disk structure (and
by extension the X-ray light curve) than to alter the 
optical emission from the disk.

Equation (\ref{eq:ascale}) is not compatible with the original suggestion by
Balbus and Hawley (1991) that the dynamo process should
saturate in equipartition between the magnetic field and
the gas pressure.  This suggestion was motivated by the notion
that the dynamo process arises from the random stretching of
field lines in a turbulent medium (\cite{b50}, \cite{k67}).
On the other hand, this kind of dynamo is rigorously justifiable only
in the limit where magnetic forces are negligible, which is
obviously not the case here.  Equation (\ref{eq:ascale})
may be compatible with the
notion that the instability gives rise to an incoherent
dynamo effect (\cite{vb97}), although
this predicts a somewhat steeper dependence on geometry
as $(c_s/r\Omega)\rightarrow 0$. However, the saturation
of the instability as the magnetic field approaches equipartition
may explain the discrepancy (\cite{v99}), and is, in any case,
necessary to explain the low sensitivity to geometry seen 
in the numerical simulations (whose parameters are generally
equivalent to $c_s/r$ of order unity).  We stress that since the dependence is
fundamentally one of geometry rather than Mach number, the current
generation of numerical simulations are compatible with either
model.  Here we will explore the consequences of equation (\ref{eq:ascale})
for vertical structure,  although we will also discuss the alternative,
that $\alpha$ is constant down to a limiting value of the magnetic
Reynolds number.

Another major issue in vertical structure calculations is
the role of convection.  Although convection is unlikely
to account for any significant level of radial angular momentum
transport ($\propto \langle v_r v_{\theta}\rangle$, where $v_r$ and
$v_{\theta}$ are the fluctuating velocity fields)
in Keplerian disks (\cite{bh98}), it may still play an important
role in vertical energy transport ($\propto \langle v_z \delta T\rangle$,
where $\delta T$ is the fluctuating temperature field).
The use of standard mixing length theory (MLT) is even more uncertain
in disks than in stars, and previous structure calculations have
included simple nonlinear closure schemes (\cite{ch86}, \cite{cchp87},
\cite{can92}), detailed linear analyses of the convective modes
(\cite{rpl88}), and a reduced effective mixing length (\cite{can92}).
Some of these calculations included the
assumption that angular momentum transport is mediated by
convective motions, so that the results cannot be extended to
ionized disks.  There is an additional problem with the use
of nonlinear closure schemes since such schemes typically fail
to reproduce experimental and numerical results in extremely
strong shearing environments, although there has been some progress
in the intervening years (cf. Speziale et al. 1991; Speziale and
Gatski 1997).  A particular problem with such schemes is that they
tend to underestimate the degree of anisotropy for low
Rossby numbers.  In any case, none of the
previous work included the effect of diffusive damping and
mixing due to turbulent motions acting across the reduced radial
length scales.   How can we determine
the `typical mode' of convection (i.e. in the spirit of MLT)
when differential rotation
and BH turbulence are simultaneously present?
To account for nonlinear effects within the context of
linear theory, we start with the linear perturbation
theory of convection for Keplerian disks and find the
modes most resistant to the dissipative effects of 
shear and turbulent mixing.  We use the
properties of these linear modes
as a guide to the nonlinear state and 
the typical velocity and eddy size of convective cells.
We defer discussion of the empirical basis for this approach
to the last section of this paper.

The purpose of the paper is to calculate the vertical
structure and  thermal equilibrium S-curves for a variety
of astrophysical disks, including the effects of turbulent
energy transport throughout the whole thickness of the disk, 
reasonable modifications of MLT for
convection in disks, and the scaling formula for $\alpha$
given in equation (\ref{eq:ascale}).
Standard MLT and the associated vertical energy transport
normally used in Keplerian disk models are reviewed
in \S 2. Section 3 describes the turbulent energy
flux driven by the BH instability. This energy 
flux is particularly important if convection is suppressed.
We examine the conditions for the suppression of convection
in \S 4. In \S 5, we propose a new model for vertical energy transport
in disks.  We present and discuss our results
for DNe, SXTs, and AGNs in \S 6. 
Finally, we summarize our results and present our conclusions in
\S 7.

\section{Standard MLT with a scaling law for $\alpha$}

The basic equations for the vertical
structure of a geometrically thin,
Keplerian, optically thick disk are derived from
mass conservation, radiative and convective energy transport, 
and thermal and hydrostatic equilibrium.
These give four differential
equations for the column density $\Sigma$,
the total vertical energy flux $F$, the total
pressure $P$, and the temperature $T$ as a function of
distance from the disk midplane:
(\cite{mm82}, \cite{cc88})
\begin{eqnarray}
{d\Sigma \over dz}&=&2\rho=2{P_g\mu \over RT},\label{df1} \\
{dF\over dz}&=&{3\over 2}\alpha \Omega P,\label{df2} \\
{dP\over dz}&=&-\rho g_z=-\rho \Omega^2 z=-{P_g\mu \over RT}
\Omega^2 z,\label{df3} \\
{dT\over dz}&=&-{3\kappa \rho \over 4acT^3}F_{rad}
=-{3\kappa \left( {P_g\mu \over RT}\right) \over
4acT^3}\left( {\nabla \over \nabla_r}\right) F,\label{df4}
\end{eqnarray}
where $z$ is the vertical coordinate perpendicular to
the disk plane and is zero at the mid-plane,
$\mu$ is the mean molecular weight, $\kappa$ is
the Rosseland opacity, $F_{rad}$ is the radiative flux,
$\nabla$ and  $\nabla_r$ are defined as $d\ln T/d\ln P$
associated with the radiative flux $F_{rad}$ and the total
flux $F$ respectively (\cite{cox68}):
\begin{eqnarray}
F_{rad}&=&{4ac\over 3}{T^4\over \kappa \rho \lambda_p}
\nabla. \\
F&=&{4ac\over 3}{T^4\over \kappa \rho \lambda_p}\nabla_r.
\end{eqnarray}
The gas pressure $P_g$, is the difference of the total
pressure and radiation pressure; i.e.,
\begin{equation}
P_g=P-{1\over 3}aT^4.
\end{equation}
We note that $\nabla_r$, the gradient $d\ln T/d \ln P$ when the total
energy transport is carried by radiation, does not necessarily correspond
to any real gradient.  It is a 
mathematical convenience used to calculate the total flux $F$. 
In our model the turbulent energy flux is always nonzero, although
sometimes negative, and
the total flux is never carried by radiation alone.
We will address the nature of the turbulent energy transport later.

These four differential equations have four boundary conditions.
The definition
of the column density $\Sigma=2\int \rho\,dz$ gives 
$\Sigma(z=0)=0$. The reflection symmetry about
the mid-plane gives 
$F(z=0)=0$. The other two boundary conditions follow from the
definition of the photosphere,
$P(z=h)={2\over 3}g_z(z=h)/\kappa (z=h)$ and
$F(z=h)=\sigma T^4(z=h)$, where $h$ is the half-thickness
(from the mid-plane to the bottom of the photosphere) of
the disk.

We need one more differential equation for the variable $\alpha$.
Since we lack a model for the vertical distribution of dissipation,
we will assume that
the scaling law for $\alpha$
depends only on the mid-plane sound speed. This
gives us a trivial differential equation 
\begin{equation}
{d\alpha \over dz }=0,\label{df5}
\end{equation}
with the boundary condition
\begin{equation}
\alpha(z=0)=50\left( {c_s(z=0)\over \Omega r}\right)^{1.5},
\end{equation}
where
\begin{equation}
c_s^2(z=0)={P_{mid}\over \rho_{mid}}=
{P_{mid}\over \left( P_{mid}-{1\over 3}aT_{mid}^4 \right)
{\mu_{mid} \over RT_{mid}}}.
\end{equation}
There are good reasons to think that equation (\ref{df5}) is not
correct (see, for example, figure 10 in \cite{SHGB96}), and
will have to be modified once we understand the $z$ dependence of
the dynamo process.

Now we turn to the cooling processes which play important
roles in disks in the relevant temperature range, i.e.
thousands to tens of thousands of degrees K.
The usual approximations
for vertical energy transport in a Keplerian,
optically thick accretion disk
include the Rosseland approximation for the radiative
energy flux $F_{rad}$ and MLT for the convective
energy flux $F_{conv,MLT}$. Hence,
the total vertical flux in the convectively unstable region
is given by
\begin{equation}
F_{rad}+F_{conv,MLT}
={4ac \over 3}{T^4 \over \kappa \rho \lambda_p}\nabla
+ {1\over 2}\rho v_{MLT} \Lambda c_p T {1\over \lambda_p}
\left( \nabla -\nabla' \right), \label{frfc}
\end{equation}
where $v_{MLT}$ is the vertical speed of the
convective bubble in the mixing-length theory,
$\Lambda$ is the
mixing length for convection, $c_p$ is the
specific heat at constant pressure and
$\nabla'$ is $d \ln T/d \ln P$ estimated inside the
convective bubble. Notice that using
$\nabla'$ instead of the adiabatic gradient
$\nabla_{ad}$ implies heat
exchange between a convective bubble and its
environment over the course of its lifetime. This
tends to erase entropy inhomogeneities
and reduce the convective flux. In general,
this process is described by the equation
\begin{equation}
{\Gamma' \over 1-\eta' }={ \nabla - \nabla' \over \nabla'
-\nabla_{ad}},\label{Gamma'}
\end{equation}
where $\Gamma'$, the convective efficiency, is given by 
\begin{equation}
\Gamma' = {c_p\kappa \rho^2
v_{MLT} \Lambda \over 8acT^3 }.\label{isogamma'}
\label{mlt1}
\end{equation}
Here we have assumed isotropic convection so that
the volume to surface ratio is $V/{\cal A}=\Lambda/6$.
The convective speed in this expression is
$v_{MLT}=-i\Lambda \bar N/ 2\sqrt{2}$, the square of
Brunt-V\"ais\"al\"a frequency with radiative
damping ${\bar N}^2=-g_zQ(\nabla - \nabla') /\lambda_p$,
and $\eta'$ is the ratio of the excess heat generated
within the convective bubbles to the excess energy radiated
by the convective bubbles during their lifetime, and $Q$
is the coefficient of expansion.
Here the energy exchange could be the radiative
losses given by $\Gamma'$, or it could include
any other diffusive effect given by $\eta'$.
We shall see that this is
one of the crucial points in this paper. 
Equation (\ref{frfc}) can be
rewritten in terms of $\nabla_r$ and the convective
efficiency for convective bubbles $\Gamma'$.
\begin{equation}
\nabla_r = \nabla + a_0 \Gamma' (\nabla - \nabla'),
\label{conv}
\end{equation}
where $a_0=3$. 
Equations (\ref{Gamma'}) and (\ref{conv})
were originally used for stellar models and
their use for disks is somewhat questionable.
The important feature of traditional MLT is
implicit in our expression for $\Gamma'$ which describes
nearly isotropic convection without any drag
on the motion of convective bubbles\footnote{To be more
precise, traditional MLT assumes that the turbulent 
drag is due only to the interaction of convective bubbles
with each other, and can be represented by an efficiency for
the conversion of the free energy into the kinetic energy
of the bubbles.}.
We shall see below
that equation (\ref{Gamma'}) should be modified
to take into account the unique properties of accretion
disks: shear as a result of differential
rotation and turbulent mixing, and turbulent drag, due to the 
BH instability.

\section{Turbulent Heat Flux Driven by the BH Instability}

As discussed above, it seems reasonable to assume that
viscous heating and angular
momentum transport in hot accretion disks are driven by
the BH instability.
Just as convective mixing leads to vertical energy
transport through mixing, turbulent eddies
driven by BH instability will also produce
vertical energy transport at rate proportional
to the local entropy gradient.   
Hereafter the term ``turbulent flux'' will be used to
denote the energy flux induced by the BH instability.
As noted before, in principle the turbulent thermal
conductivity is not necessarily equal to the turbulent
viscosity  (\cite{rud}).  However, given the approximately
isotropic nature of the BH instability, and for the
reasons given in the discussion of equations (\ref{eq:vert})
and (\ref{conduct}) we will take $\chi_T=\nu$.
By analogy with the formula for the convective
flux, we can obtain the turbulent flux by
replacing  the convective diffusion coefficient
$v_{MLT}\Lambda$ with $\nu$. 
In other words, the turbulent flux
driven by BH stability is given by
\begin{equation}
F_{turb}={1\over 2}\rho \nu c_p T{1\over \lambda_p}
(\nabla - \nabla'' ),\label{turbflux}
\end{equation}
where $\nabla''$ is ($d\ln T/d\ln P$)
estimated inside a turbulent bubble driven by
the BH instability. Introducing $\nabla''$ 
means that we also assume that there is an exchange
of radiative energy between turbulent eddies driven
by BH instability and their surroundings. We
can define a turbulent efficiency $\Gamma''$ for
turbulent eddies analogous to $\Gamma'$ for isotropic
convection. In other words, replacing the term
($v_{MLT}\Lambda$) with
$\nu$ in the expression for $\Gamma'$, we have
\begin{equation}
\Gamma''={c_p \over 8ac}{k\rho^2 \nu \over T^3},
\label{dpdef}
\end{equation}
and its relation with the entropy inhomogeneities
associated with turbulent cells is analogous to
that for convective cells:
\begin{equation}
{\Gamma'' \over 1-\eta''}={\nabla - \nabla'' \over
\nabla'' - \nabla_{ad}},\label{Gamma''}
\end{equation}
where  $\eta''$ is the ratio of the excess heat generated within
the turbulent bubbles to the excess heat radiated from the bubbles
during their lifetime. A similar set of 
equations for the radiative damping of
the turbulent energy
flux due to shear instabilities
in a differentially rotating star has been studied
by Maeder (1995).  In physical terms,
the radiative efficiency is the ratio of the radiative cooling
time scale to the eddy turnover time.
The cooling time scale due to radiation in the optically
thick limit is the square of
the radiative diffusion length associated
with the turbulent bubbles divided by the radiative
conductivity $4acT^3/(3\kappa \rho^2 c_p)$.

There are several reasons to believe that
turbulent energy transport could be very important.
First, convection will vanish when entropy per mass
increases away from the midplane,
while the turbulent flux driven by BH instability
is always present.  Moreover, the turbulent flux can carry
energy toward hot mid-plane when a disk is convectively
stable. Second,
the turbulent flux inferred by equation (\ref{turbflux})
has its maximal value when $\nabla'$ is equal to
$\nabla_{ad}$. In this particular case of weak radiative losses,
equation (\ref{turbflux}) is roughly equal to
\begin{equation}
{P \nu \over \lambda_p}(\nabla -\nabla_{ad})
\sim {P \nu \over \lambda_s},
\end{equation}
where $\lambda_s$ is the entropy scale height.
The ratio of this flux to the radiative
flux ($\sim \dot M \Omega^2$) is $\sim \lambda_p
/\lambda_s$, which will be of order unity unless a disk
is nearly adiabatic (\cite{vis93}). In 
other words, the turbulent energy flux can be as large as
the total energy flux when radiative losses are small.
Finally, and most importantly, convection
can suffer from strong shearing in a Keplerian disk and from 
strong mixing in a turbulent disk. We expect that
turbulent mixing will be important whenever 
convective mixing is suppressed in spite of a unstable
entropy gradient. We address this point in the
next section.

\section{The Suppression of Weak Convection}

Modeling convection is often considered one of the main
sources of uncertainty in stellar models.
Despite this MLT has provided a reasonable framework for
achieving an approximate understanding of stellar interiors 
and has been broadly applied to accretion disks. 
However, the question of its applicability to accretion
disks has to be raised since strong shearing and
mixing can be expected to have a dramatic effect 
on the disk environment, and to vitiate the usual
assumption of nearly isotropic convection.
Shearing will shorten the radial width of 
convective cells whenever convection is
weak (i.e. the Brunt-V\"ais\"al\"a frequency $-iN<<\Omega$).
Turbulent mixing may prevent the formation of the
kind of organized convective cells assumed in MLT.
Here we will show how to modify
MLT to take these important effects into account.

\subsection{The radial deformation of convective cells}

We begin by considering the linear perturbation equations
for nearly incompressible (Boussinesq), axisymmetric convection in
an accretion disk some local viscosity.  They are
\begin{eqnarray}
(i\bar\omega+k^2\nu)v_r +ik_r \Psi - 2\Omega v_{\theta}&=&0, \label{per2}\\
(i\bar\omega+k^2\nu)v_{\theta}+{\Omega \over 2}v_r &=&0, \label{per3}\\
(i\bar\omega+k^2\nu)v_z + ik_z\Psi +\delta g_z &=&0,\label{per4}\\
i\bar\omega\delta-v_z \partial_z \ln\left({P^{1/\Gamma}\over\rho}\right)
&=&0,
\label{per5}\\
k_rv_r+k_zv_z&=&0,\label{per6}
\end{eqnarray}
where $\bar \omega$ is the frequency measured by a local
observer rotating with the disk, $\delta\equiv \delta \rho
/\rho$ is the local fractional density perturbation,
$\Psi\equiv \delta P/\rho$ is the pressure perturbation
divided by the density, $v_r$, $v_\theta$, and $v_z$ are
velocity perturbations, $k_z$ and $k_r$ are the vertical
and radial wavenumbers, and the
vertical gravitational
field is given by $g_z=z\Omega^2+2\pi G \Sigma(z)$, where the
second term is from the self-gravity of the disk. 
Self-gravity is
not negligible for AGN disks for small $\alpha$
(\cite{sc81}, \cite{ls86}, \cite{can92})
which, as we shall see later, can happen when
we use equation (\ref{eq:ascale}).  This system of equations
implicitly assumes that the vertical wavenumber is small compared
to the pressure scale height, and other vertical scale lengths.
Of course, this is wrong, but it makes only a small difference
if we use these equations and identify $k_z$ with the inverse
pressure scale height at the end.

A more serious issue is that we have simply ignored the inevitable
loss of axisymmetry in the convective turbulence.
Again, this is not as bad as it first appears.  The
existence of strong shear
will tend to suppress high angular wavenumber modes.
Physically this follows from the fact that modes with
short angular wavelengths will respond to strong shearing
by deforming into a spiral wave with a short radial wavelength.
This will have an overwhelming effect on mode structure when
the shearing time is shorter than the growth time of
an unstable mode.  Complete suppression of the instability
follows when the radial wavelength becomes longer than
the characteristic scale for the variation of the comoving
frequency. In other words, we require
\begin{equation}
k_\theta<< k_r {\gamma\over\Omega}, \label{ktheta}
\end{equation}
where $\gamma$ is the convective growth rate measured by
a local observer rotating with the disk, and $k_\theta$
is the azimuthal wavenumber.
When $k_\theta$ satisfies this constraint,
the effects of shearing will be modest.  In general, this
will make $k_\theta$ no larger than $k_z$.  When we calculate
the volume to surface ratio for a
typical convective cell shape below we will assume that
$k_\theta\sim k_z$.

Finally, we have left the effects of turbulent damping out of
the entropy equation (equation (\ref{per5}).  We do this to
avoid double counting, since we add this effect to the 
usual effect of radiative damping when calculating the 
efficiency of convection.

We can get a rough understanding of how the local shear
modifies the shape of convective cells by considering
the case $\nu=0$.  Then equations (\ref{per2}) through
(\ref{per6}) imply
\begin{equation}
\left( {\Gamma_c^2-\gamma^2 \over \Omega^2 +\gamma^2} \right)
=A,
\end{equation}
where $A\equiv k_z^2/k_r^2$,
$\Gamma_c\equiv -iN=-ig_zQ(\nabla-\nabla_{ad})/\lambda_p$, 
and $\gamma\equiv -i\bar \omega$
is the convective growth rate. If convection is weak in
the sense that $\Gamma_c <<\Omega$, $A<<1$ and the radial
length of convection is shortened by the strong shear.
In fact, we have $A\sim (\Gamma_c/\Omega)^2$ if the dominant 
mode has a growth rate which is some fraction of 
$\Gamma_c$. 

The radial thinning of convective cells under the influence
of the local shear has important consequences for the suppression of
convection. First, this radial distortion makes
it much easier for turbulent diffusion, due to the BH instability,
to damp out convective cells.
The damping rate due to radial mixing becomes
$\nu k_z^2/A\sim \nu \Lambda^{-2}/A$ which is stronger than
the isotropic result, $\nu /\Lambda^2 \sim \alpha \Omega$
(assuming that $\Lambda \sim \lambda_p$).  Using our
full perturbation equations (\ref{per2}) - (\ref{per6}) we
obtain the dispersion relation
\begin{equation}
{\gamma \over \Omega} \left( 1 + \left({\gamma \over
\Omega}+{k^2\nu \over \Omega}\right)^2 \right) A
=\left( \left( {\Gamma_c \over \Omega}
\right)^2 -{\gamma \over \Omega}{\gamma+k^2 \nu \over
\Omega}\right) \left( {\gamma \over \Omega}+
{k^2\nu \over \Omega}\right).\label{dispersion}
\end{equation}
We see that a sharp increase in $k^2/k_z^2$ due to the effects
of shear is associated with much smaller values of $\gamma$.

The second effect associated with the eddy
distortion is that the average radiative diffusion length is
now reduced.
This means that the expression of radiative
efficiency equation (\ref{isogamma'}) needs to be modified
for radially deformed convective cells. The definition
of the radiative efficiency gives (\cite{cox68})
\begin{equation}
\Gamma'={3\kappa \rho^2 c_p v_z \over 2acT^3 \Lambda}
\left( {V\over {\cal A}} \right) l_T,
\end{equation}
where $l_T$ is the distance between
the center of the convective element, where the
temperature perturbation is a maximum, and its surface. 
The ratio $V/{\cal A}$ is the volume to surface ratio
for a convective bubble.  Under normal circumstances
$l_T$ is nearly proportional to $V/{\cal A}$ and is
$\Lambda/2$ for the isotropic case. We can
make a plausible estimate for this length scale by
comparison with $V/{\cal A}=\Lambda/6$ in the isotropic case.
We use $l_T=3(V/{\cal A})$ 
for an anisotropic parallelepiped and assume for simplicity 
that the convective cells have a parallelepiped
shape. Then the expression for radiative efficiency becomes
\begin{equation}
\Gamma'={kc_p\rho^2 v_z \Lambda \over 8acT^3}f,
\label{anigamma'}
\end{equation}
where we have defined the anisotropy factor
\begin{equation}
f\equiv [3(V/{\cal A})/(\Lambda/2)]^2={9A\over(1+2A^{1/2})^2}.
\end{equation}
This has a maximum value of one when $A=(k_z/k_r)^2=1$, 
that is, in the isotropic limit.
Note that this reduction in convective efficiency has nothing
to do with turbulent mixing, and is purely the result of
shearing. 

Equation (\ref{anigamma'}) shows that the radiative efficiency is reduced
when the eddies are slow, $v_z< v_{MLT}$, due to the turbulent drag
and radially thin, that is $f<<1$.
We have already incorporated turbulent drag into the dispersion
relation (cf. equation (\ref{dispersion})). Radiative
damping can be added by replacing the Brunt-V\"ais\"al\"a frequency
for the adiabatic case with one that uses $\nabla'$, as discussed
in section 2.
The dimensionless gradients can be related to $\Gamma'$ 
(equation (\ref{anigamma'})) through equation (\ref{Gamma'}). Before we
do this, we need to include one last damping effect, the role of
turbulent mixing in smoothing entropy gradients between convective bubbles.
This is the topic of the next section.

\subsection{Mixing losses}

The diffusion, and eventual loss, of the entropy inhomogeneities
associated with convective bubbles due to turbulent mixing
is similar to the effects of radiative loss
(\cite{tz97}, \cite{DAlessio})
and can therefore be treated in an analogous fashion. 
We do this by extending equation (\ref{Gamma'}) which
describes efficiency of convection in the face of
radiation transfer between entropy inhomogeneities.
The diffusive term $\eta'$ in equation (\ref{Gamma'})
can have either sign, depending on the physical context.
For example, convection in extremely hot stellar interiors
can involve a local competition between local thermonuclear
energy generation and neutrino losses (\cite{cox68}). 
In a similar way, entropy transfer through turbulent mixing
in accretion discs can be modeled by defining $\eta'$
as the ratio of the turbulent to radiative heat transfer 
between convective bubbles. The turbulent mixing
loss under pressure equilibrium is equal to the turbulent
conduction $(1/2)\rho \nu c_p \Delta T$, where $\Delta T$
is the temperature gradient within the convective bubbles.
The radiative loss is $4acT^3 \Delta T/(3\kappa \rho)$
in the optically thick limit. The ratio between these two,
using definition of $\Gamma''$ given in equation (\ref{dpdef}), gives
\begin{equation}
\eta'=-a_0 \Gamma'',
\end{equation}
which gives a decrease in convective efficiency.

In summary, shearing and turbulence act together to reduce
convective heat transport.
Shear alone shortens the radiative diffusion length,
turbulent mixing alone smooths away the entropy inhomogeneities
associated with convective bubbles, and the interplay of
these two gives a turbulent drag on convective eddies.
We can account for these three factors by
modifying equations (\ref{Gamma'}) and (\ref{mlt1}) using 
equation (\ref{anigamma'}) to obtain
\begin{equation}
{\Gamma' \over 1-\eta'}
={c_p \over 8ac}{k \rho^2 v_z \Lambda \over T^3}f
\left( {1\over 1+a_0 \Gamma'' }\right)
={\nabla - \nabla' \over \nabla' - \nabla_{ad}}.
\label{Gamma'_nabla}
\end{equation}
When these corrections lead to a suppression of convection,
the turbulent heat flux induced by BH instability can have
a significant effect on the vertical structure of
accretion disks. As we shall see below,
weak convection, where the convective growth rate is less
than $\Omega$, is typically suppressed in a wide variety
of accretion disks.  However, before we can show this 
we need to determine `typical' convective eddy sizes and speeds
allowing for shearing and turbulent mixing.  We do so
in the next section.
 
\section{A New Model for Vertical Energy Transport}

A full treatment of vertical energy transport needs to
include radiation, convection, and turbulent mixing
driven by BH instability.  As discussed above,
MLT needs to be corrected by incorporating the effects of
shear and turbulent mixing. The starting point for
our corrections is the linear dispersion relation equation 
(\ref{dispersion}) for convection. However, it would be 
unreasonable to expect the use of purely linear perturbation 
equations to give us an accurate model of the fully nonlinear 
end state.  The linear dispersion relation
gives a family of solutions, any one of which is an
equally valid starting point, but whose properties can
vary tremendously. This means that we can give physically
reasonable estimates for the properties of the nonlinear 
regime from the linear theory only if we are
able to find a `typical' linear mode whose properties
are the best guide to the nonlinear state; i.e., the
mode which is most resistant to viscous damping and
secondary instabilities. We are not going to propose a
model for the turbulent spectrum for convection.
Instead, our objective is to produce
estimates for the typical vertical velocity $v_z$ and the
typical radial extent of a convective cell (i.e. still
one-mode model), that reduce
to the usual mixing length estimates when convection
is strong and small scale turbulence is weak, but which
incorporate physics neglected by MLT.

The growth of the convective instability is ultimately limited
by dissipative turbulence generated from secondary instabilities
and from the interactions between convective modes.  
We can quantify this by considering the ratio of the mode shear
to the growth rate squared, or
\begin{equation}
P_s\equiv{S^2\over\gamma^2},
\end{equation}
where
\begin{equation}
S_{ij}={1\over2}(v_{i,j}-v_{j,i}),
\end{equation}
and $\gamma$ is the mode growth rate.  When this is large
the mode will quickly succumb to micro instabilities, so
by minimizing $P_s$ we can choose the linear mode most
relevant to the nonlinear limit.  This criterion is not, by itself,
adequate because we are also dealing with the effects of
turbulent mixing on small scales, which will introduce
an effective viscosity into the system.  This suggests that
we also need to minimize the ratio of viscous damping to
the mode growth rate.  Since this is an independent effect,
we will add the two in quadrature by defining a second ratio
\begin{equation}
P_\nu\equiv{(k^2\nu)^2\over\gamma^2}
\end{equation}
and minimize
\begin{equation}
P_\nu+P_s.
\label{criterion}
\end{equation}
The relative weighting of these two terms is not obvious, but
an equal  weighting seems the most reasonable choice.
Note that when the system is heavily damped, 
$P_\nu$ will be large for all modes, and minimizing the sum will
reduce to a minimization of $P_\nu$.  This guarantees that
this criterion will always pick the modes that can still grow
as we approach the limit of marginal stability. 

For an axisymmetric, incompressible convective roll, we have
\begin{equation}
4S^2=k^2v_\theta^2+{k^4\over k_r^2}v_z^2.
\end{equation}
We can combine this with
the expression for $v_{\theta}$ from
equation (\ref{per3}), replacing $i\bar \omega$
with $\gamma$, to obtain
\begin{equation}
P_s=\left( {(1+A)\Omega^2 \over 16 (\gamma+k^2 \nu)^2}+
{(1+A)^2\over 4A} \right) \left( {k_z v_z \over \gamma}
\right)^2.\label{P_s}
\end{equation}
The last factor of this equation should be approximately
independent of mode shape ($A$).  The dispersion relation 
equation (\ref{dispersion}) in
the inviscid, non-shearing, and isotropic limit gives
$\gamma=\Gamma_c/\sqrt{2}$. Since we need to reproduce MLT
in this limit we have
\begin{equation}
v_z=\sqrt{2}v_{MLT}{\gamma \over \Gamma_c}={\Lambda \gamma/2}.
\label{v}
\end{equation}
In MLT we take $k_z=2/\Lambda$ in MLT, that is, the perturbation
wavelength is half the size of an isotropic eddy.  Consequently,
the last factor in equation (\ref{P_s})
is equal to one. In the absence of rotation and
viscosity, minimizing $P_s$ implies $k_r^2=k_z^2$
or $\Lambda_r=\Lambda$.  In this limit $P_s$ does not depend
on the viscosity, and the influence of viscosity on mode shape
is felt entirely through $P_\nu$.

We note that our treatment assumes that the vertical scale
length $k_z^{-1}$ remains the same as in standard MLT, 
$\Lambda/2$.  This follows from our use of the minimization
principle.  Both $P_s$ and $P_\nu$ involve wavenumbers
through the ratio of $k_r$ to $k_z$, which does not
affect the value of $k_z$, and through the strength of viscous
effects, which always favor a small $k_z$.  Therefore
we assume that $k_z^{-1}$ is as small as possible, 
about one pressure scale height.

With this in mind, we need to minimize
\begin{equation}
P_{tot}=\left( {(1+A)\Omega^2 \over 16(\gamma +k^2\nu)^2}
+{(1+A)^2\over 4A} \right)+{(k^2 \nu)^2\over \gamma^2},
\end{equation}
as a function of $A$ and $\gamma$, with
\begin{equation}
k^2\nu={8\over 3}\alpha \Omega {1+A \over A}
{\lambda_p g_z \over (\Lambda \Omega)^2}.\label{drag}
\end{equation}
This minimization is subject to constraints
from the dispersion
relation equation (\ref{dispersion}) and to the total energy transport
equations involving radiation, convection, and turbulent mixing
driven by BH instability.
The purpose of the paper is to solve the
differential equations (\ref{df1}), (\ref{df2}),
(\ref{df3}), (\ref{df4}), and (\ref{df5})
with the corresponding boundary conditions
to investigate the vertical structures
and, furthermore, to
see how the disk structure changes when one
uses the scaling law for $\alpha$.

To solve these  equations,
we need to express the variable
$\nabla$ (or say $F_{rad}$)
in the equation (\ref{df4}) in terms
of the thermodynamical variables we want
to solve. Since the formulae for $\nabla$ depend upon
the vertical energy transport mechanisms, the thermal gradient
$\nabla$ has different forms within and outside 
convection zones. Furthermore, the usual criterion
for convective instability by comparing $\nabla_r$
with $\nabla_{ad}$ needs to be modified since
$\nabla_r$ is never physically relevant, due to the 
constant presence of turbulent flux. Therefore,
we adopt a convective criterion
involving the turbulent flux, where
the thermal gradient $\nabla_r$
is replaced with $\nabla_{RC}$ which is $(d\ln T/d\ln P)$
of the medium when convection is absent (\cite{DAlessio}).
Using this new criterion, we are ready to solve for
$\nabla$ inside and outside the convection zones.

We begin with the case $\nabla_{RC} > \nabla_{ad}$.
Then all three energy transport mechanisms coexist within the
convectively unstable region. Hence
the total vertical energy flux, $F$, is
\begin{eqnarray}
F&=&F_{rad}+F_{conv,damp}+F_{turb}\\
&=&{4ac \over 3}{T^4 \over \kappa \rho \lambda_p}\nabla
+ {1\over 2}\rho v_z \Lambda c_p T {1\over \lambda_p}
\left( \nabla -\nabla' \right)
+ {1\over 2}\rho \nu c_p T {1\over \lambda_p}
\left( \nabla -\nabla'' \right),\nonumber \label{turb}
\end{eqnarray}
Combining this equation with the definition of $\nabla_r$
gives
\begin{equation}
\nabla_r =\nabla + a_0 \Gamma' {1\over f}
\left( \nabla -\nabla' \right) 
+a_0 \Gamma'' \left( \nabla -\nabla'' \right), \label{Ftot}
\end{equation}
with the definition of $\Gamma'$ in equation (\ref{anigamma'})
and the definition of $v_z$ is equation (\ref{v}).
Moreover, the convective efficiency $\Gamma'$ and the
turbulent efficiency $\Gamma''$ can be related to
the $\nabla$s by using equations (\ref{Gamma'_nabla})
and (\ref{Gamma''}). Since the
heat generation by viscosity is proportional to
$\alpha P$, which is a constant inside and outside
the convective bubbles, $\eta''$ should be zero
and $\eta'=-a_0 \Gamma''$ as discussed above.

Combining equations (\ref{Gamma''}), (\ref{anigamma'}), 
(\ref{Gamma'_nabla}), (\ref{v}) and (\ref{Ftot}) 
we can find the thermal gradients in
terms of $\Gamma'$, $\Gamma''$, $f$ and therefore in terms of
$\gamma$ and $A$.  We obtain
\begin{equation}
\nabla = (1-\zeta )\nabla_r +\zeta \nabla_{ad},\label{nabla}
\end{equation}
where
\begin{equation}
1-\zeta \equiv \left( 1+{a_0{\Gamma'}^2 f \over
\Gamma' + 1+a_0\Gamma''}+{a_0{\Gamma''}^2 \over \Gamma''+1}
\right)^{-1}.
\end{equation}
Note that $\zeta$ has lost its original meaning referring
to convective efficiency (\cite{cox68}) and used here purely
for convenience.
Equation (\ref{nabla}) is one constraint to the minimization
problem
arising from the total energy transport. Using it
we can write $\Gamma_c$ as
\begin{eqnarray}
\Gamma_c^2&=&{g_z Q\over \lambda_p }(\nabla-\nabla')\\
&=&{g_z Q\over \lambda_p}{\Gamma' \over \Gamma' +
1+a_0 \Gamma''} (1-\zeta)(\nabla_r - \nabla_{ad}).\nonumber
\end{eqnarray}
Now we can combine all the constraints together by substituting 
this expression for $\Gamma_c^2$
into the dispersion relation, so that the result
becomes the only constraint
on the minimization problem in terms of the two variables
$\gamma$ and $A$. Once the minimization is solved numerically, 
the value of $\zeta$ is determined by
$\gamma$ and $A$. This in turn gives
$\nabla$ in terms of $\nabla_r$ and $\nabla_{ad}$
using equation (\ref{nabla}).
We get the ratio of the radiative flux to the total
flux ($F_{rad}/F=\nabla /\nabla_r$) in the differential
equation (\ref{df4})
and then solve the
four differential equations.

Outside the convection zone, $\nabla_{RC} < \nabla_{ad}$,
we have
\begin{eqnarray}
\nabla_r &=&\nabla +a_0 \Gamma'' \left( \nabla -
\nabla'' \right), \label{frft} \\
\Gamma'' &=&{\nabla -\nabla'' \over \nabla'' -
\nabla_{ad} },\\
\Gamma''&=&{c_p \over 6ac}{\kappa \rho^2 \nu \over T^3}.
\end{eqnarray}
Solving for $\nabla$, we have (\cite{mae96})
\begin{equation}
\nabla ={1\over 1+a_0 \Gamma'' -{a_0 \Gamma''
\over \Gamma'' +1}}\left( \nabla_r +
{a_0 {\Gamma''}^2 \over \Gamma'' +1}\nabla_{ad}\right).
\label{delfrft}
\end{equation}
We can see the physical meaning of the turbulent efficiency
in this case (\cite{mae96}). 
When $\Gamma''$ goes to infinity, we have
\begin{equation}
\nabla \to \nabla_{ad},\qquad \nabla'' \to \nabla_{ad},
\qquad {F_{turb}\over F}={\nabla_r -\nabla \over \nabla_r}
\to {\nabla_r -\nabla_{ad} \over \nabla_r}.
\end{equation}
Note that when turbulent transport is very efficient, it
does not necessarily imply that the turbulent flux carries 
of the flux. 
When $\Gamma''$ goes to zero, we have
\begin{equation}
\nabla \to \nabla_r,\qquad \nabla'' \to \nabla_r,
\qquad {F_{turb}\over F}={\nabla_r -\nabla \over \nabla_r}
\to {\nabla_r -\nabla_r \over \nabla_r}=0.
\end{equation}
Finally, $\nabla$ in equation (\ref{frft}) gives the
expression of $\nabla_{RC}$:
\begin{equation}
\nabla_{RC}={\nabla_r + a_0 \Gamma'' \nabla'' \over
1+a_0\Gamma'' }.
\end{equation}
$\nabla_{RC}$ is equal to $\nabla$ in equation
(\ref{delfrft}) outside the convection zones. We note 
that using $\nabla_{RC}$ instead of $\nabla_r$ does not
matter much in the end. When they differ, it will be
because of either strong turbulent mixing or weak convection.
In either case turbulent mixing will tend to
suppress convection altogether.

\section{Results and Discussion}

The computer program used to solve the vertical structure equations
is a modification of the one described in two previous papers
(\cite{cw84}, \cite{cc88}). The adiabatic gradient
$\nabla_{ad}$, mean molecular weight $\mu$, and coefficient of
expansion $Q=-\partial \ln \rho /\partial \ln T$ were supplied by
H. Saio for these papers using the input physics given in \cite{ibe63}.
We assumed a solar abundance of 
Na, K, Fe (assuming that
the overall ionization potential $\chi_M=5.76$ eV), and C.
We use the Rosseland opacities $\kappa$
of \cite{cs69} for
$\log T(K)>3.9$, those of \cite{ajr83} for
$\log T(K)<3.9$, and those of \cite{pmc85} for $T(K)<1220$.
We smoothed $\log\kappa$ linearly
over $3.8<\log T(K)<4.0$ and $1220<T(K)<1675$
to eliminate the slight
discontinuity in those temperature ranges as done in the
original code. In the calculation, we set the mixing
length $\Lambda$ for convection equal to
max($\lambda_P$, $z$). The nonlinear minimization problem for
the subjective function $P_{tot}$ subject to the
equality constraint equation (\ref{dispersion}) with positive
variables $\gamma$ and $A$ is achieved by the Fortran
routines DONPL2 (\cite{spe98}) and NLPQL (\cite{sch85}).

\begin{figure}
\epsscale{.8}
\plotone{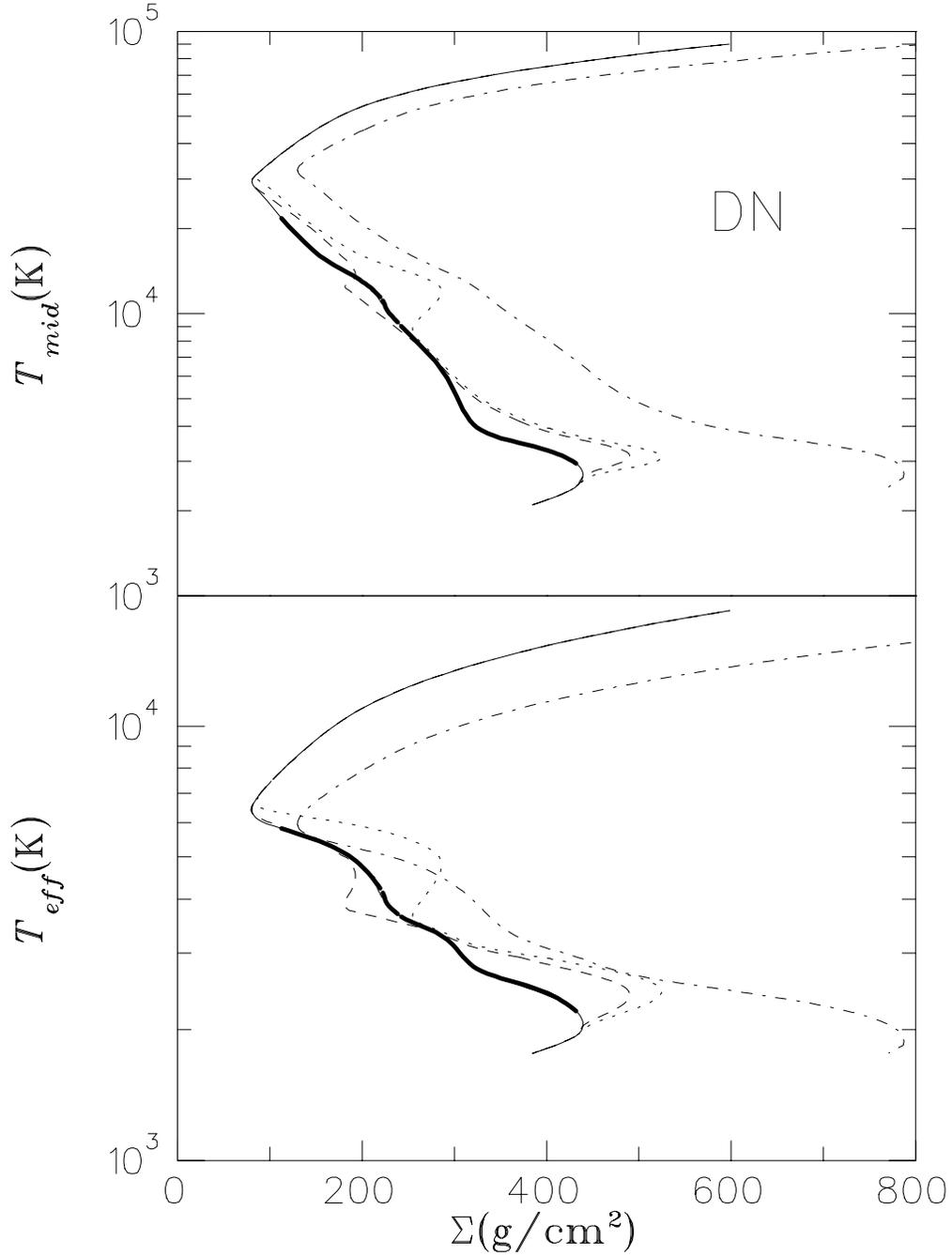}
\caption{S-curves of DN disks
($M_{WD}=1\,\msun$, $\log r\,{\rm cm}=10.5$) for different vertical
energy transport models: $F_{all,damp}$ (thicker solid line),
$F_{all,nodamp}$ (dotted line), $F_{rad}+F_{conv,MLT}$
(dashed line), and $F_{rad}+F_{turb}$ (solid line). 
While these curves are based on the scaling law
$\alpha=50(c_s/r/\Omega)^{3/2}$, the dash-dotted line denotes
the S-curve for $F_{rad}+F_{turb}$ based on the scaling
law $\alpha=20(c_s/r/\Omega)^{3/2}$.\label{fig1}}
\end{figure}

Fig.~\ref{fig1}
shows the thermal equilibrium S-curves for a
DN at $r=10^{10.5}$ cm with a white
dwarf mass $=1\ \msun$.
Every point in the thermal equilibrium curves is one solution
to the differential equations (\ref{df1}) to (\ref{df4}) and
each curve is associated with a
different model for the vertical energy transport.  We do
this to illustrate the effect of the different mechanisms
and to show how our model differs from the previous standard
model.  Here we show the results of a model with heat carried
by radiation and MLT convection
($F_{rad}+F_{conv,MLT}$), by radiation and turbulent flux driven
by BH instability ($F_{rad}+F_{turb}$), by all three
transport mechanisms but with MLT convection ($F_{tot,nodamp}$), or
by all three transport mechanisms with modified (damped) 
MLT ($F_{tot,damp}$).  We have also added a fifth curve 
showing the sensitivity of our results to the choice of $\alpha_0$.

These curves differ mainly on the middle, unstable, branch 
because the disk is partially ionized at these temperatures and
the resulting opacity peak lowers the efficiency of radiative
transfer relative to convective or turbulent mixing.
The surface densities
at the left and right turnovers of the unstable branches 
are usually denoted by $\Sigma_{min}$ and $\Sigma_{max}$.
In this case, only $\Sigma_{max}$ shows any sensitivity to
the input physics.  The short stable branch within the 
unstable branch, which appears in some of the curves, is an
effect of very strong mixing.
It has been previously shown, for convective
mixing, that the unstable branch moves up and to the right in
the $T-\Sigma$ plane as the mixing length $\Lambda$ increases
(\cite{mo83}, \cite{poj86}, \cite{lmr94}). Increasing the mixing
length is equivalent to enhancing the strength of convection.
This suggests that regardless of the mixing mechanism at work,
the upper-right shift and the intermediate branch of the unstable branch
indicate the presence of stronger mixing.  This explains why
both the standard MLT convection model and the addition of
turbulent mixing without damping produce qualitatively similar
curves, with the latter lying consistently to the right.
On the other hand, the curve for radiation and turbulent mixing
alone lies on top of the curve with damped convection.  This
joint curve lies consistently to the left at lower temperatures,
although the old standard model ($F_{rad}+F_{conv,MLT}$)
crosses it near the intermediate branch.  The fact that the
addition of damped convection produces no significant changes
indicates that turbulent damping is very efficient in this case.
The tendency of the joint curve to lie below, and to the left,
of the othe curves shows that vertical heat transport via
turbulent mixing is generally less effective than undamped
convective transport.  The difference between the joint curve
and the old standard model largely vanishes at high temperatures
because the value of $\alpha$, and the consequent strength of
turbulent mixing, increases in case.  When convection is
very strong the convective growth rate becomes of order $\Omega$
and
\begin{equation}
{F_{conv}\over F_{turb}}\sim {\Gamma_c \over \alpha \Omega}
\sim {1\over \alpha},\label{conv/turb}
\end{equation}
so that a small $\alpha$ implies a small role for turbulent mixing.

Finally, we note that the results of our model differs from previous work in two
interesting ways.  Firstly,
the intermediate stable branch is completely
suppressed and this
implies that DNe accretion disks do not come to rest, even
briefly, midway through the ionization transition. This result
would weaken the ``stagnation effect" (\cite{min88})
which relied partly on the existence of this middle unstable branch
to account for UV delay in the early rise of DN outbursts. 
One would still have stagnation,
however, due to the large specific heat at the temperatures
between $\sim10^4$ and $2\times 10^4$ K.
Secondly, the value of $\Sigma_{max}$ is reduced by about 20\%
and this is
quantitatively interesting. However, we see from this figure that this
correction is less important than arriving at a definitive value of
$\alpha_0$ in the scaling law.  Choosing $\alpha_0=20$, which is
a conceivable, albeit extreme, choice, can almost 
double $\Sigma_{max}$ while producing only a minor change in $\Sigma_{min}$.

\begin{figure}
\plotone{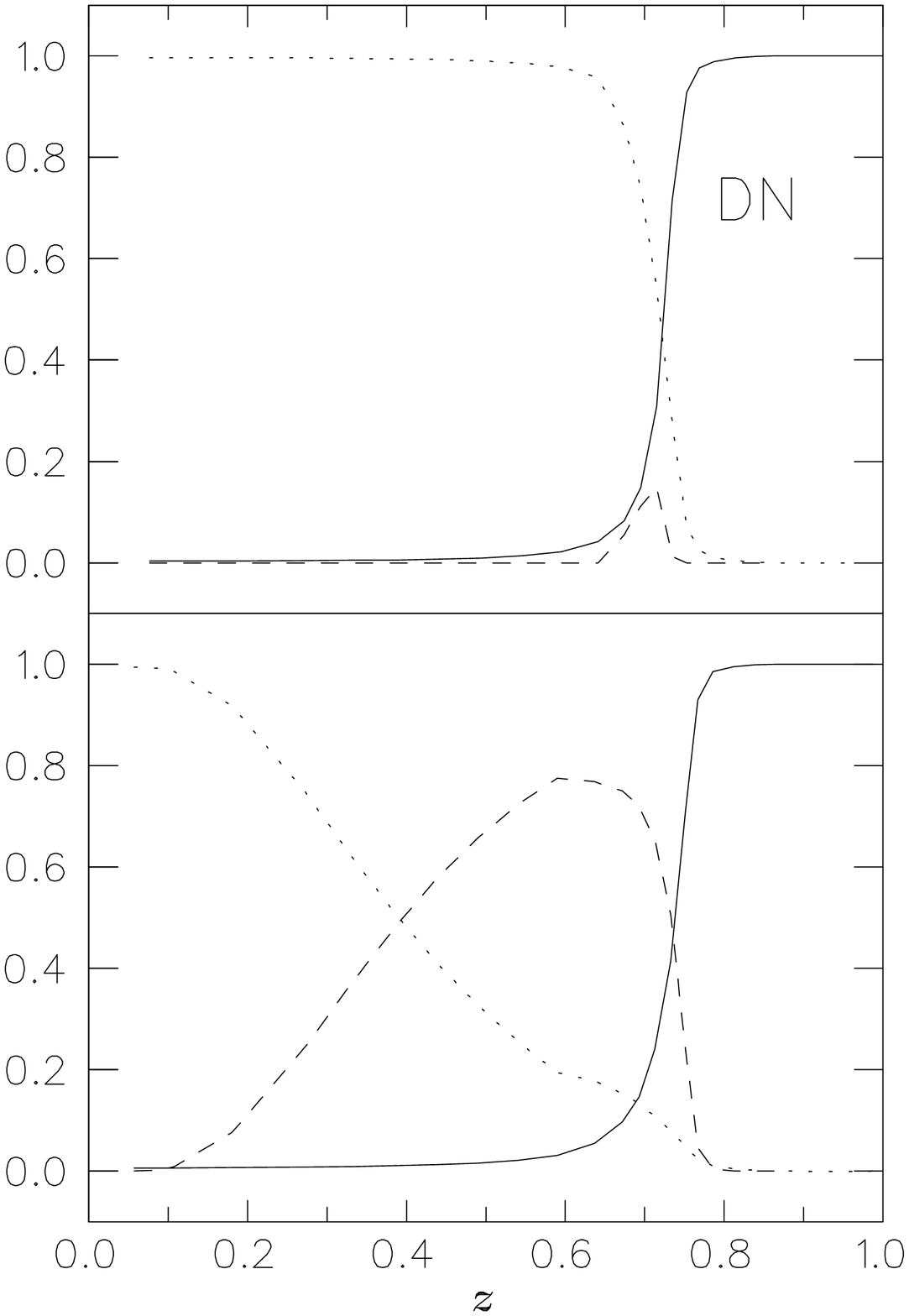}
\caption{Fractional
energy fluxes for DN disks
($M_{WD}=1\,\msun$, $\log r\,{\rm cm}=10.5$) along the
vertical direction $z$ for the same
half-thickness $\log h\,{\rm cm}=8.80$ in the cases of
damped convection (top panel) and MLT convection (bottom
panel). $F_{rad}/F_{total}$, $F_{conv}/F_{total}$, and
$F_{turb}/F_{total}$ are denoted by a solid line, a dashed line,
and a dotted line respectively.\label{fig2}}
\end{figure}

Fig.~\ref{fig2} shows the fractional heat fluxes as a function
of height within a DN accretion disk.
The top panel shows a model with turbulent transport and 
damped convection while the bottom 
one is a model with undamped convection.  
These two cases have the same total
half-thickness $\log\,h({\rm cm})=8.80$ and lie on the middle
branch of the S-curve.  The undamped convection model is roughly
at the upper end of the intermediate branch.  Both models have
$\log\,r({\rm cm})=10.5$ and use the scaling
law for $\alpha$. There is no significant change in $F_{rad}$ between
the two models, indicating that the total mixing flux is
roughly the same, but in the case of damped convection this
flux is almost entirely carried by $F_{turb}$.
We note that formally the convectively unstable region 
($\nabla_{RC}>\nabla_{ad}$) is not dramatically smaller in the
damped case.  Instead, while both models have broad, and roughly equal,
convectively unstable zones ($\approx h$), the shrunken convective
zone shown in Fig.~\ref{fig2} in the damped case is due to the
lack of solutions to the minimization problem for weak convection.
{\it This result for weak convection shows that 
the presence of an available source of free energy is not a guarantee
of instability if the environment supplies strong damping}. 
In Table~\ref{tb-DN_conv} we see that where convection is strongest
in the undamped convection model, it just barely survives in the
damped model, even though the eddies are almost isotropic
and roughly adiabatic ($\Gamma'>1$).  At this level, even their
complete elimination would not change the S-curve significantly
since the curve for $F_{all,damp}$ already coincides with the 
convectionless model.  The factor $1/(1+a_0 \Gamma'')$
describing the turbulent losses is smaller as $\Gamma'$
increases. This means that convective and turbulent mixing
becomes stronger (or weaker) than radiation simultaneously
and this in turn explains why turbulent mixing is not important
on high and low temperature stable branches of the S-curves 
where radiative transport dominates and the convectively unstable 
region is narrow.  Although $F_{turb}$ exists in the convectively stable 
region and carries energy toward the mid-plane, increasing the
temperature gradient,
$F_{turb}$ behaves in other respects like $F_{conv}$ with a 
shorter mixing length and is severely
damped by radiative losses when radiative transport is important.
Our calculations show that $\Gamma''\sim 10^{-2}$ in the stable hot
and cold states for the $F_{rad}+F_{turb}$ curve in Fig.~\ref{fig1},
rendering mixing transport of any kind largely irrelevant. This
is why all the S-curves coalesce in the hot and cold states.
Fig.~\ref{fig2} also shows that for both models turbulent mixing
dominates near the midplane where the disk is more opaque and
where convection is weak due to small $g_z$.
The numbers in parentheses in
Table~\ref{tb-DN_conv} (or in
Table~\ref{tb-LMXB_conv}, and Table~\ref{tb-AGN_conv}) are the
corresponding results for the case $F_{all,nodamp}$ for the same
$h$. As expected, these numbers are 
larger than those for damped convection. The growth rate, $\gamma$, 
for undamped (MLT) convection in these tables
is $\Gamma_c /\sqrt{2}$ as defined in the preceding
section. 

Table~\ref{tb-DN_alpha} shows the values of $\alpha$ along
the S-curves in Fig.~\ref{fig1}.  The outburst duration
for DNe seems to imply that $\alpha\sim 0.2$ in the
hot state (cf. \cite{s84}, \cite{can94}).
Evidently the scaling law leads to a considerable
overestimate (up to about 0.6) as we go up the hot state
branch of the S-curve for DN disks.  In addition, 
the use of the scaling law for $\alpha$ in the hot state in 
DN disks leads to a contradiction with the Bailey
relation which says that the decay time associated with
the decline from outburst in DNe increases with orbital period.
Both of these points suggest that the scaling law should
start to saturate as $\alpha$ rises above $\sim 0.1$.
We note that the scaling law is motivated by the behavior
of SXTs, which have no Bailey relation and much stronger
evidence for an exponential decline from outburst.
One simple compromise might be to let $\alpha$ rise
to a maximum value of $0.2$ and then abruptly go over
to a constant.  This would have no effect on
the middle and lower branches of the S-curve, where turbulent
mixing and the suppression of convection can have some effect.
However, we note that a more physically motivated geometrical
dependence for $\alpha$ involves a prolonged transition
from a steep scaling to essentially constant behavior
(\cite{v99}).  An alternative possibility is to choose
a smaller value of $\alpha_0$, equivalent to asserting that
the central black hole mass in SXT's is smaller.
Table~\ref{tb-DN_alpha} also displays
the values of $\alpha$ based on the formula $20(c_s/r/\Omega)^{3/2}$
along the S-curve indicated by the dash-dotted line in
Fig.~\ref{fig1}, equivalent to asserting that the central
mass in SXT systems is typically $\sim 4 M_{\sun}$, which is
a very low value.  

\begin{figure}
\plotone{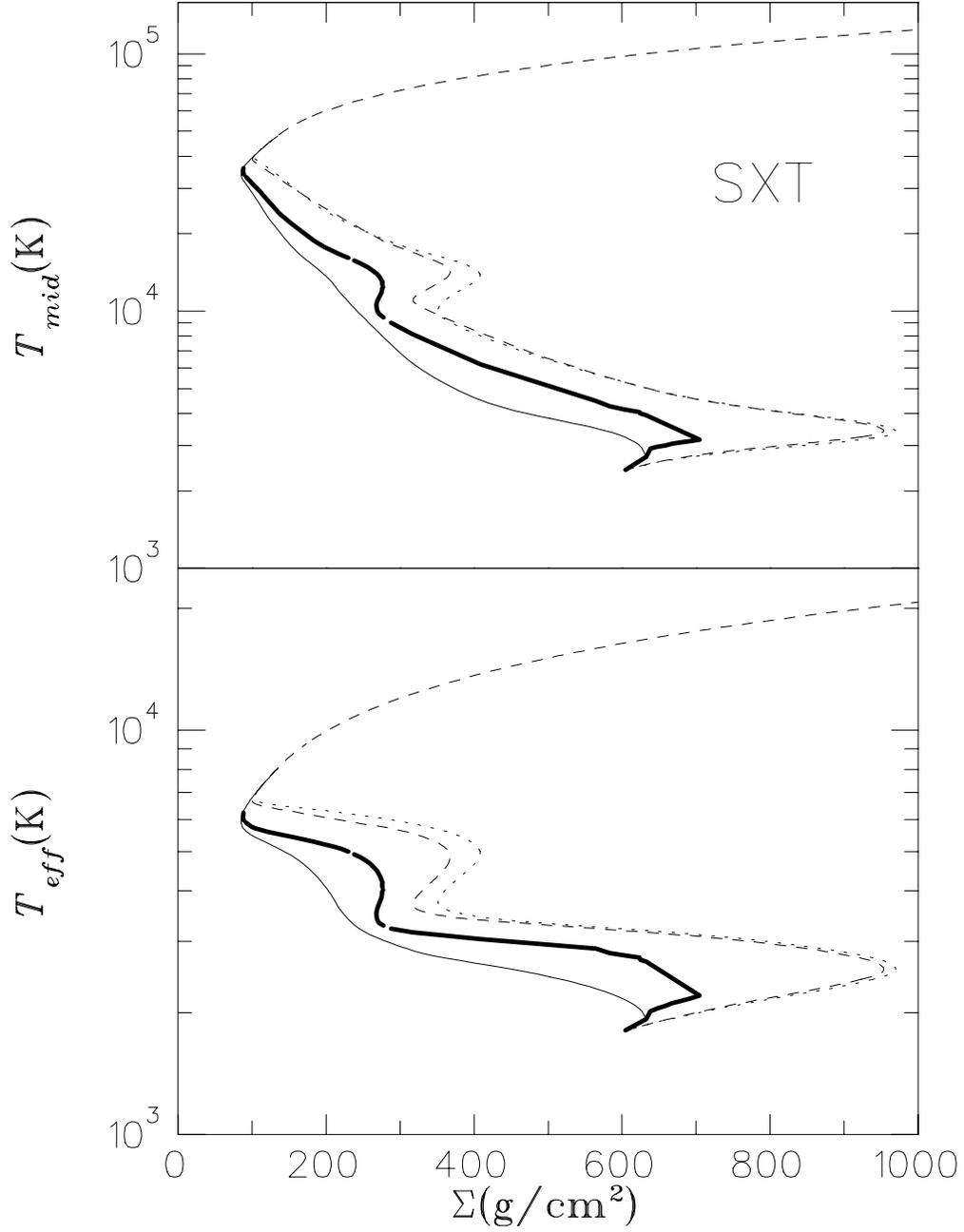}
\caption{S-curves of SXT disks
($M_{WD}=10\,\msun$, $\log r\,{\rm cm}=10.5$) for different vertical
energy transport models: $F_{all,damp}$ (thicker solid line),
$F_{all,nodamp}$ (dotted line), $F_{rad}+F_{conv,MLT}$
(dashed line), and $F_{rad}+F_{turb}$ (solid line). 
All curves are based on the scaling law
$\alpha=50(c_s/r/\Omega)^{3/2}$.\label{fig3}}
\end{figure}

The S-curves for SXT disks are illustrated in Fig.~\ref{fig3}
at $r=10^{10.5}$ cm with the central object of $10\ \msun$.
The much larger $\Sigma_{min}$ and $\Sigma_{max}$ along with
the more pronounced hysteresis curve for the total flux
($F_{rad}+F_{conv,MLT}$)
reflect on stronger convection in SXTs
than that in DNe at $\log\,r\,({\rm cm})=10.5$. 
This results from the larger $\Sigma$ for SXTs.
The energy flux radiated from the surface is given by
\begin{equation}
\dot M \Omega^2 \sim \nu \Sigma \Omega^2 \sim \alpha c_s^2 \Sigma \Omega.
\end{equation}
Since S-curves for different systems
operate at the similar $T_{eff}$, $\kappa$ (i.e. the condition
for partially ionized hydrogen) and 
$c_s^2 \propto T_{mid} \propto \tau^{1/4} \propto \Sigma^{1/4}$,
the above equation implies
\begin{equation}
\Sigma \propto \left( {1\over \alpha \Omega }\right)^{4/5}.\label{Sigma1}
\end{equation}
Moreover, the scaling law for $\alpha$ gives
\begin{equation}
\alpha \propto \left( {r\over M} \right)^{3/4} \Sigma^{3/16},\label{alpha}
\end{equation}
which actually does not depend strongly on $\Sigma$.
Together with equation (\ref{Sigma1}) this gives
\begin{equation}
\Sigma \propto M^{4/23} r^{12/23},\label{Sigma}
\end{equation}
in accordance with the usual result that the column
densities in the ionization transition region drop as a
cooling wave propagates inward in a disk (for recent
calculations see \cite{ccl95}, \cite{hmd98}).  Similarly,
we can obtain a crude estimate of the strength of convection
on the unstable branch from equation (\ref{isogamma'}).  Since
$\kappa$ is roughly the same for disks with similar ionization
fractions and temperatures, and since we have already seen
that the midplane temperature differences will be slight, we
can rewrite equation (\ref{isogamma'}) as
\begin{equation}
F_{conv,MLT}/F_{rad} \sim \Gamma' \propto \Sigma^2\Omega{v_{MLT}\over c_s} \propto
{M^{39/46}\over r^{21/46}},\label{conv/rad}
\end{equation}
where we have ignored the change in $v_{MLT}/c_s$.
This implies that convection is relatively more important for larger
central masses and at smaller radii.  In fact, SXT disks have 
larger column densities than DN disks for ionization transitions
at the same $r$ and consequently stronger mixing.
Fig.~\ref{fig3} shows that
the curves ($F_{rad}+F_{conv,MLT}$) 
and ($F_{rad}+F_{turb}$) are more different than in 
Fig.~\ref{fig1}.  Conversely, the smaller difference between
the curves $F_{all,nodamp}$ and ($F_{rad}+F_{conv}$) also
illustrates the increased importance of convection (and reduced
importance of turbulent mixing).  Adding to this effect
is the smaller $\alpha$ which reduces the importance of
turbulent mixing and damping.
The end result is that the damped $F_{conv}$ in SXTs is
stronger than in DNe.  Although the $F_{all,damp}$ curve
is still close to the curve with no convection, it consistently
lies to the left and includes a very small intermediate branch.
As before the value of $\Sigma_{max}$ is reduced, in this case
by about 30\%, relative to the old standard model.  Following
Table~\ref{tb-LMXB9_alpha}) we note
that the values of $\alpha$ used for these models never reach
the unrealistically high values seen on the hot branch in DNe.

\begin{figure}
\plotone{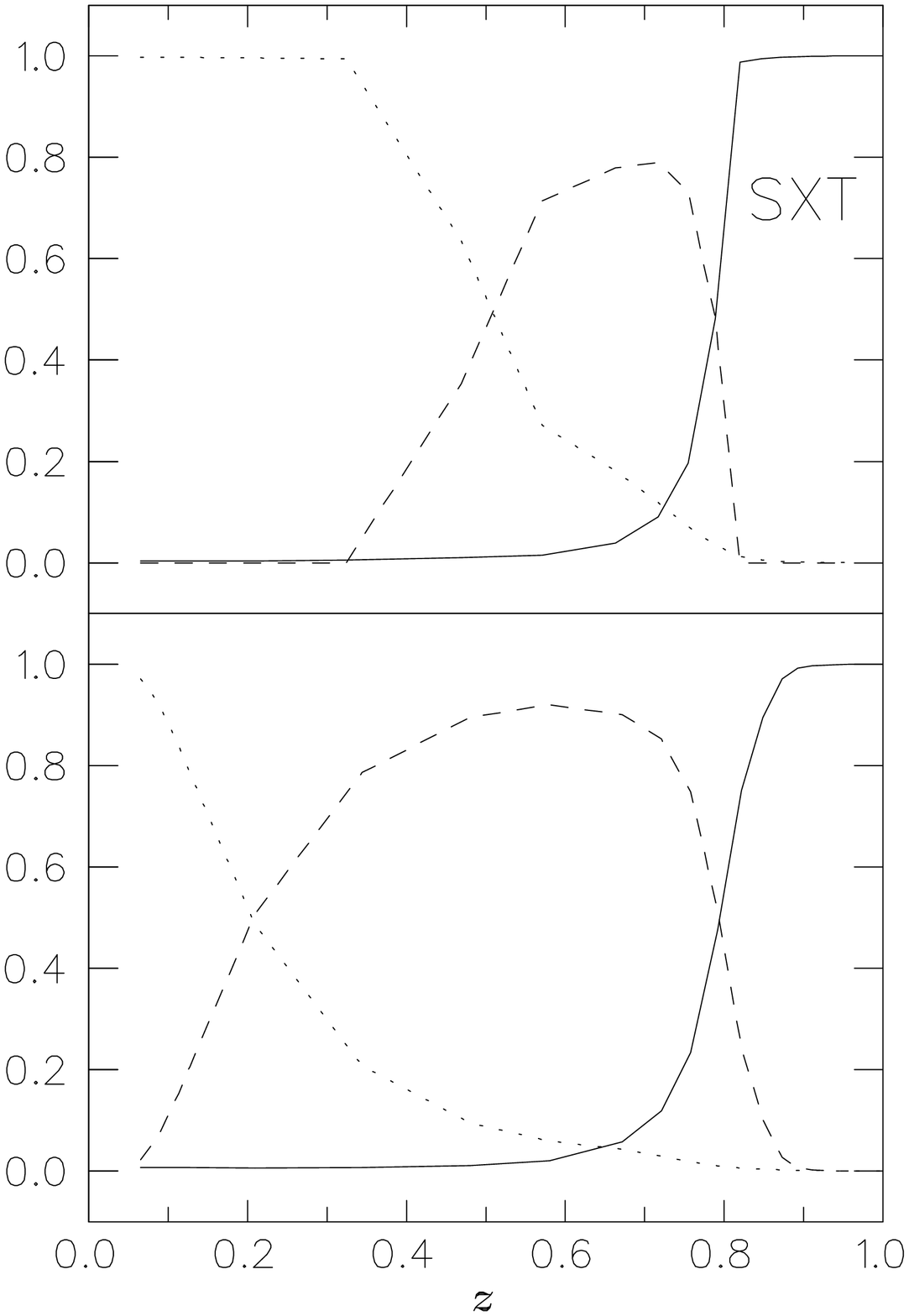}
\caption{Fractional
energy fluxes of SXT disks
($M_{WD}=10\,\msun$, $\log r\,{\rm cm}=10.5$) along the
vertical direction $z$ for the same
half-thickness $\log h\,{\rm cm}=8.33$ in the cases of
damped convection (top panel) and MLT convection (bottom
panel). $F_{rad}/F_{total}$, $F_{conv}/F_{total}$, and
$F_{turb}/F_{total}$ are denoted by a solid line, a dashed line,
and a dotted line respectively.\label{fig4}}
\end{figure}

Fig.~\ref{fig4} is corresponds to Fig.~\ref{fig2}, but for
SXTs instead of DNe.  In this case the total height 
for both models is $\log\,h\,({\rm cm})=8.33$.  The damped
convection model in the top panel lies near the upper end
of the short intermediate branch.  As expected, convection is much
more significant in SXTs in terms of both its intensity
and its vertical range. This is also seen in Table~\ref{tb-LMXB_conv} 
which shows that $P_{tot}$ is of the order of unity in the confined
convective region, suggesting that damping is weak for the strongest
convective modes.  In addition, the difference between that of damped
and undamped convective speeds is much smaller than for the
DNe cases shown in Table~\ref{tb-DN_conv}. Table~\ref{tb-DN_conv}
and Table~\ref{tb-LMXB_conv} (including Table~\ref{tb-AGN_conv}
which we will discuss later in AGN cases) also show that
for sufficiently strong convection the aspect ratio
$A$ usually increases with height since radiative damping
($\Gamma' \lesssim 1$) and turbulent drag
($k^2 \nu$, see equation (\ref{drag}))
are more severe farther from the midplane.
In fact, for other values of $h$ we find that 
slightly radially distended convective eddies
($A$ slightly larger than 1) 
can actually be favored at high altitudes.
The survival of strong convection in the middle of convectively
unstable zones and the
complete suppression of weak convection once the convective
growth rate drops raises the question of convective overshooting. 
The error that results from ignoring overshooting should be small 
in our case for two reasons.  First, the total mixing flux does not 
change if damping is added. Second, overshooting is important 
in cases where no other mixing mechanism exists.  In
accretion disks there is always turbulent
mixing.  Finally, we note that in our case convective overshooting 
still carries energy away from the mid-plane in the overshoot region
since $\nabla_{RC}>\nabla_{ad}$. This is in contrast to the
stellar case, where $F_{conv}$ is negative in overshoot region, which
is convectively stable, and this in turn increases
the temperature gradient. 

\begin{figure}
\plotone{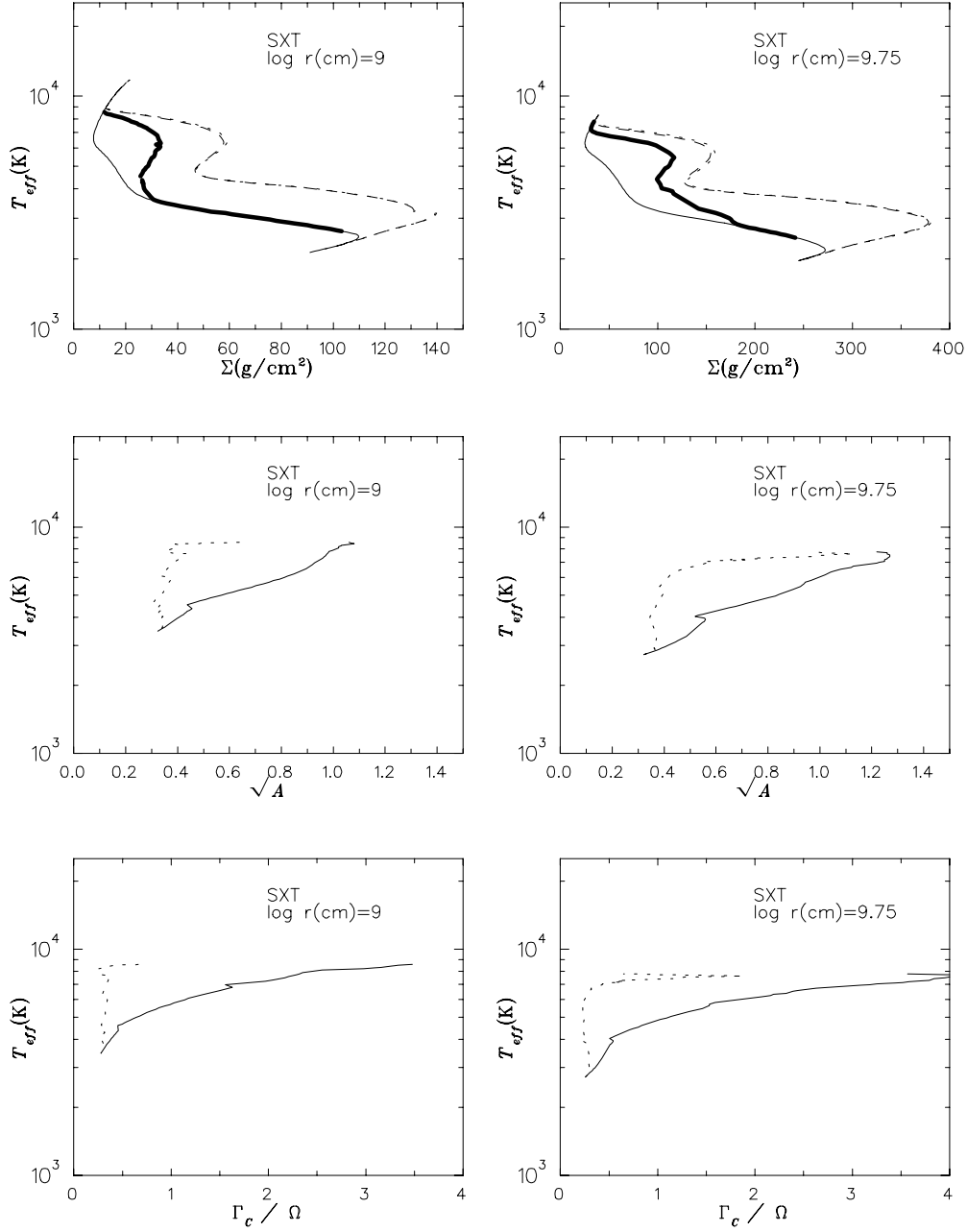}
\caption{S-curves of SXT disks around a $10\, \msun$
central black hole at $\log r\,{\rm cm}=9$ and $9.75$
for different vertical
energy transport models: $F_{all,damp}$ (thicker solid line),
$F_{all,nodamp}$ (dotted line), $F_{rad}+F_{conv,MLT}$
(dashed line), and $F_{rad}+F_{turb}$ (solid line). 
All curves are based on the scaling law
$\alpha=50(c_s/r/\Omega)^{3/2}$.\label{fig3.9}. We also
show the aspect ratio of convective cells, $A^{1/2}$ and
the strength of convection $\Gamma_c/\Omega$ in the middle
of the convective region as a function of $T_{eff}$.
The values of $A^{1/2}$ and $\Gamma_c/\Omega$ are the
maximum and minimum values for a given $T_{eff}$.}
\end{figure}

Fig.~\ref{fig3.9} shows the S-curves
for SXT disks for $\log\,r\,({\rm cm})=9$ and $9.75$. Although the
typical $\Sigma$ is smaller than at a radius of $\log\,r\,({\rm cm})=10.5$,
as expected from equation (\ref{Sigma}), the larger value of
$\Omega \Sigma^2$ (cf. equation (\ref{conv/rad}))
and the smaller value of $\alpha$ (cf. equation (\ref{alpha}) 
and Table~\ref{tb-LMXB9_alpha}) make convection slightly stronger
and turbulent mixing slightly weaker.  Consequently, the two
curves ($F_{rad}+F_{conv,MLT}$) and $F_{rad,nodamp}$ coalesce,
while we see a larger difference between the
curves for ($F_{rad}+F_{conv,MLT}$) and ($F_{rad}+F_{turb}$).
However, when we compare $\Gamma_c$ to $\Omega$, we see
that the strength of convection drops at smaller $T_{eff}$ and
that this effect is stronger at smaller radii.  The end result is that the
final curve, $F_{all,damp}$ follows the convective curves
at high temperatures and switches abruptly to the turbulent
mixing curve at low temperatures when the aspect ratio
of the convective cells drops to the point where radiative losses and
turbulent mixing turn off convection.  This leads to a pronounced
intermediate branch at small radii.  For $\log\,(r\,({\rm cm})=9$
the reduction in $\Sigma_{max}$ is less than 20\%.

\begin{figure}
\plotone{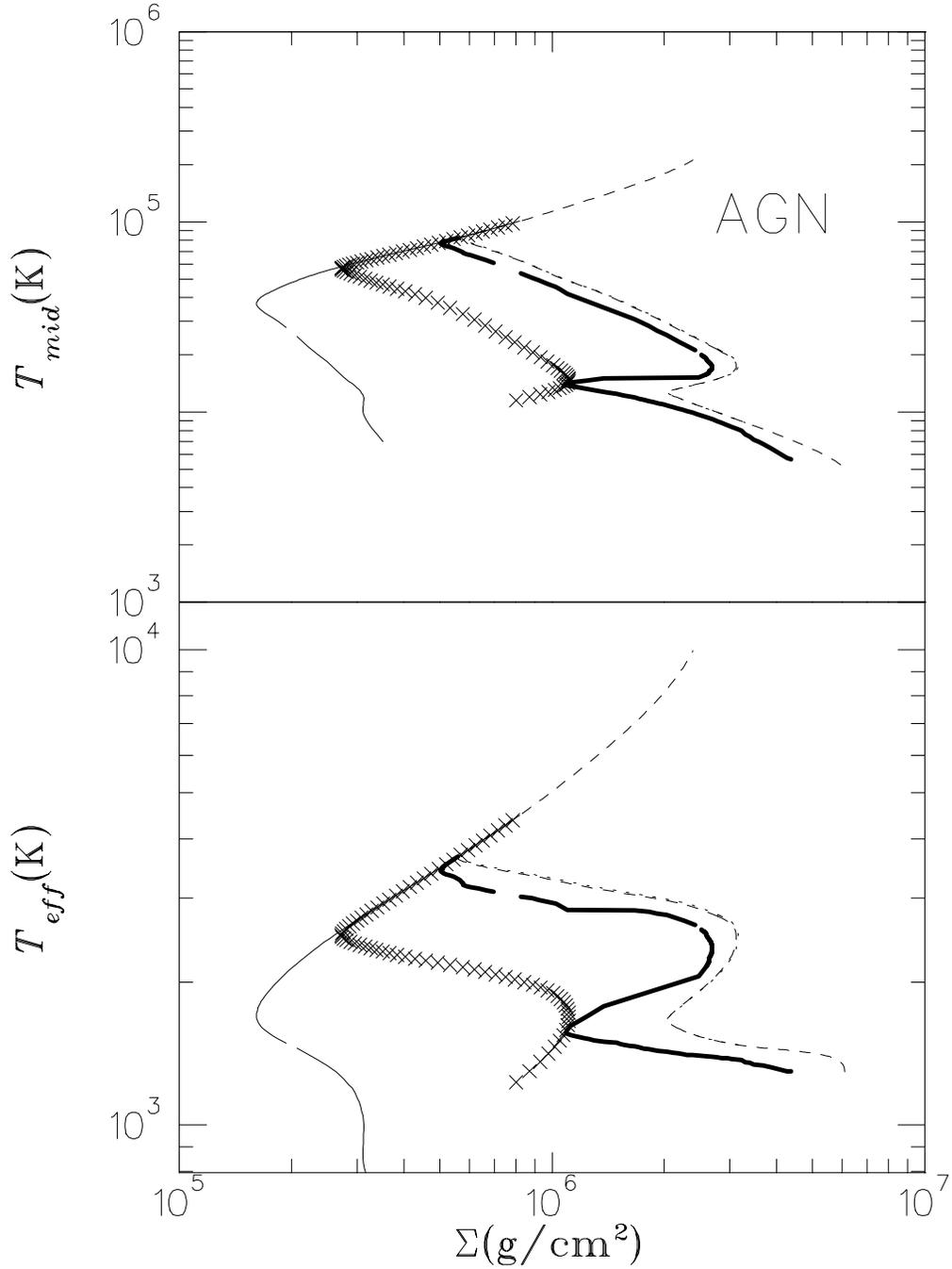}
\caption{S-curves of AGN disks
($M_{WD}=10^8\,\msun$, $\log r\,{\rm cm}=15.5$) for different vertical
energy transports: $F_{all,damp}$ (thicker solid line),
$F_{all,nodamp}$ (dotted line), $F_{rad}+F_{conv,MLT}$
(dashed line), and $F_{rad}+F_{turb}$ (solid line).
The line composed of crosses denotes the S-curve for the 
flux model $F_{rad}+F_{conv,MLT}$ with a shorter mixing-length
$\log\,(\Lambda /\lambda_p) =-0.5$.
All curves are based on the scaling law
$\alpha=50(c_s/r/\Omega)^{3/2}$.\label{fig5}}
\end{figure}

Fig.~\ref{fig5} displays the S-curves for the usual AGN parameters:
$M_{BH}=10^8\,\msun$ and $\log\,r\,({\rm cm})=15.5$.
AGN are extreme cases relative to DNe because of
their much smaller $\alpha$, which implies weaker turbulent damping
and larger $F_{conv}/F_{turb}$, and much larger $\Sigma$, 
hence larger $F_{conv}/F_{rad}$.  Consequently the
curve $F_{all,nodamp}$ lies close to the
curve ($F_{rad}+F_{conv,MLT}$) and both are far from the curve 
($F_{rad}+F_{turb}$).  There is almost no difference between the
curve $F_{all,damp}$ and the curve $F_{all,nodamp}$.
All of this is as expected for a disk where
convection is strong and turbulent mixing is weak.
The weakness of turbulent mixing is also revealed 
in the relatively large difference between $T_{mid}$ and $T_{eff}$ 
for the $F_{rad}+F_{turb}$ curves compared to the convection
S-curves; that is, the temperature difference is increased in the middle branch
as a result of inefficient turbulent mixing. 
We also ran the code for ($F_{rad}+F_{conv,MLT}$)
with a reduced mixing-length $\Lambda=0.316\lambda_p$ with the usual 
scaling law for $\alpha$. The S-curve in this case,
depicted by the line composed crosses
in Fig.~\ref{fig5}, is more or less
similar to the CCHP model, which attempts to account for shear through 
a general reduction in the mixing length, at least for
$0.01\lesssim \alpha \lesssim 1$ (\cite{can92}).  We see that
this approach exaggerates the suppression of convection for AGN. 
In our model convection remains relatively efficient for AGN despite
shearing and turbulent damping.

\begin{figure}
\plotone{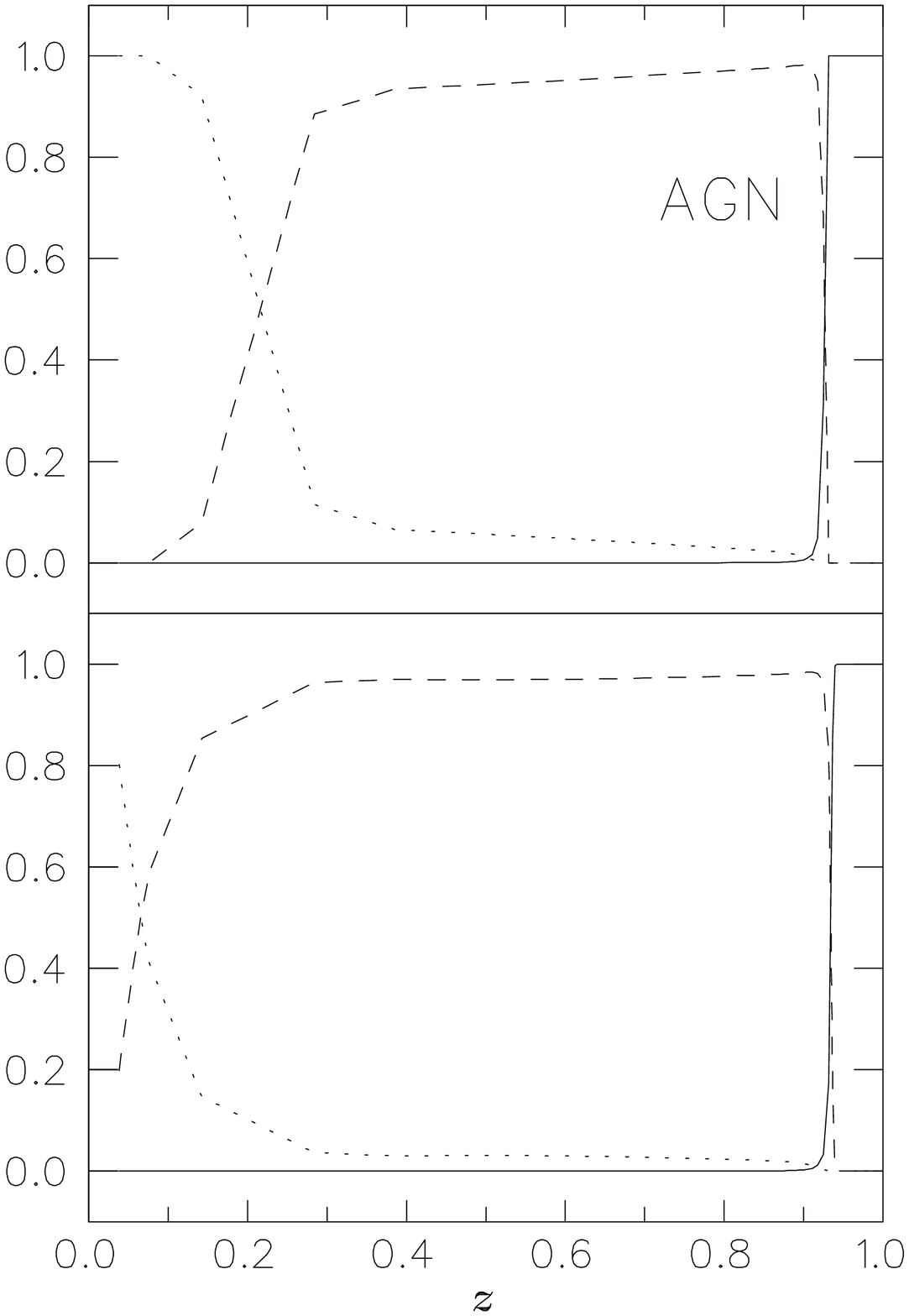}
\caption{Fractional
energy fluxes of AGN disks
($M_{WD}=10^8\,\msun$, $\log r\,{\rm cm}=15.5$) along
the vertical direction $z$ for the same
half-thickness $\log h\,{\rm cm}=12.58$ in the cases of
damped convection (top panel) and MLT convection (bottom
panel). $F_{rad}/F_{total}$, $F_{conv}/F_{total}$, and
$F_{turb}/F_{total}$ are denoted by solid line, dashed line,
and dotted line respectively.\label{fig6}}
\end{figure}

The relative importance of the vertical heat transport mechanisms
for AGN is illustrated in Fig.~\ref{fig6}.  We choose models
with $\log\,h\,({\rm cm})=12.58$ lying on the unstable branch.
We see that the overall structure of the disk is affected only
slightly by including turbulent damping of convective cells.  In
both cases turbulent mixing dominates near the midplane and decreases
rapidly in importance away from it.  Radiative transport plays
a major role only in the region just below the photosphere.
The vertical structure of the damped convection zone
is also shown in Table~\ref{tb-AGN_conv}. We see larger convective 
efficiency factors associated with radiative losses
$\Gamma'$, small turbulent drag $k^2 \nu$, and very small turbulent loss
factors $1/(1+a_0\Gamma'')$ compared to those in Table~\ref{tb-DN_conv} and
Table~\ref{tb-LMXB_conv}. As a result, $P_{tot}$ is close to unity and 
convection is well approximated, at least in terms of geometry, 
by isotropic convection without turbulent mixing.  As in the
SXT cases, we also find that $A$ is occasionally bigger than 1 
in the most robust convective modes.  The larger role
for turbulent losses $1/(1+a_0\Gamma'')$ for AGNs is really a consequence
of very small radiative losses, and does not reflect an overwhelming
suppression of convection by turbulent mixing.  We can see this
by considering the total efficiency factors obtained by dividing
the $\Gamma'$ without parentheses by the corresponding $1+a_0\Gamma''$. 
In fact, Tables~\ref{tb-DN_conv}, \ref{tb-LMXB_conv},
and \ref{tb-AGN_conv} show that the total convective efficiency
$\Gamma'/(1+a_0\Gamma'')$ for these three systems are of the same
order. This is because the small values of $\alpha$ and huge values
of $\Sigma$ for AGNs assure that radial turbulent mixing and 
radiative losses are almost negligible for them while the turbulent flux is 
extremely weak.

\begin{figure}
\plotone{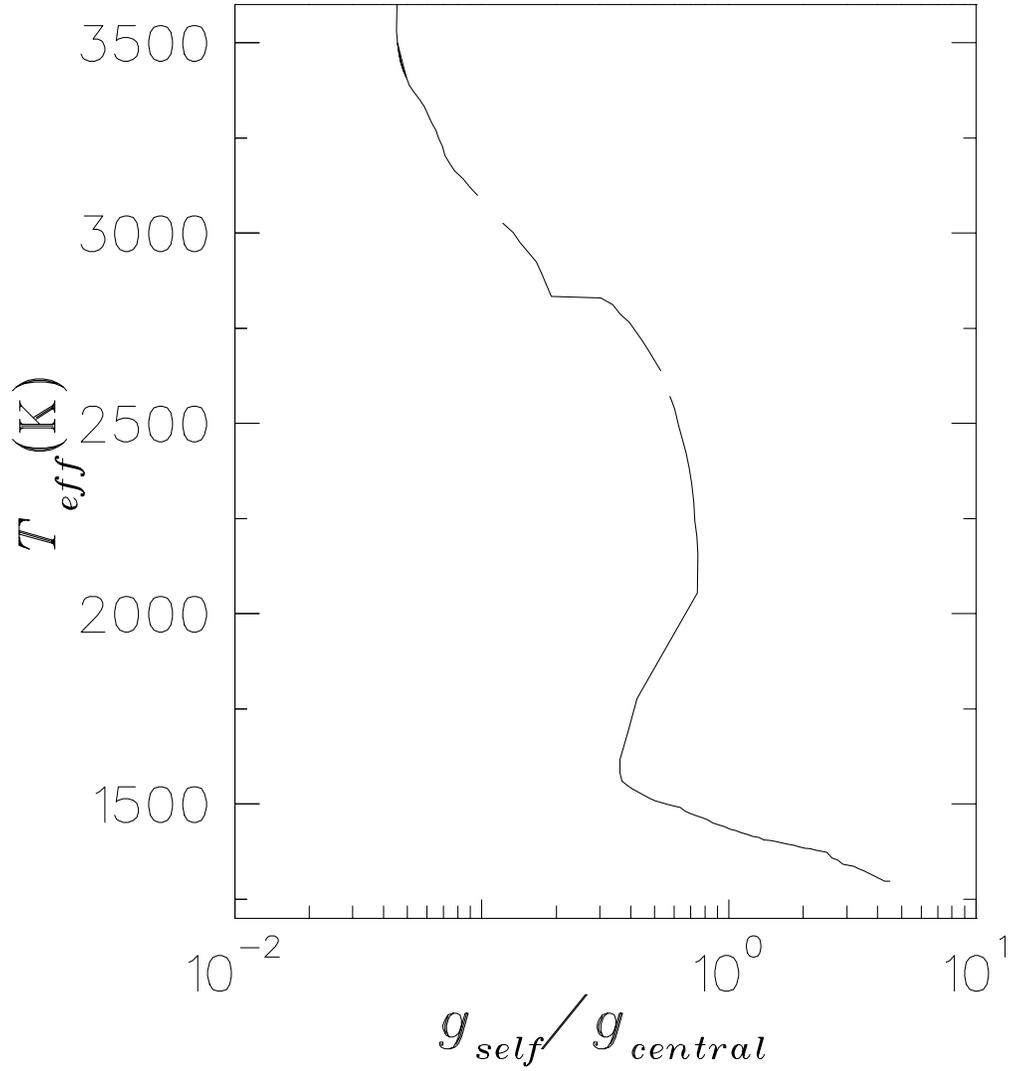}
\caption{Effective temperature $T_{eff}$
vs. the ratio of self-gravity to central object gravity
$g_{self}/g_{central}$ for AGN models whose total vertical energy
flux is $F_{all,damp}$.
\label{gself_AGN.eps}}
\end{figure}

The very large values of $\Sigma$ implied by the scaling
law for $\alpha$ applied to AGN disks (see Table~\ref{tb-AGN_alpha}) 
leads to the question of disk self-gravity.  If the self-gravity
of the disk is as important as the gravity imposed by the central
object, then local gravitational instabilities may be important and
our description of AGN disks is not self-consistent.  In
Fig.~\ref{gself_AGN.eps} we see the ratio between
self-gravity, $g_{self}=2\pi G\Sigma$, and $g_{central}=h\Omega^2$
for complete disk models of AGN with turbulent damping.
We see that the two become comparable when the AGN
disk $T_{eff}$ approaches $10^3$ K.  This should raise $\alpha$
and turn the curves sharply to the left.

\begin{figure}
\epsscale{.7}
\plotone{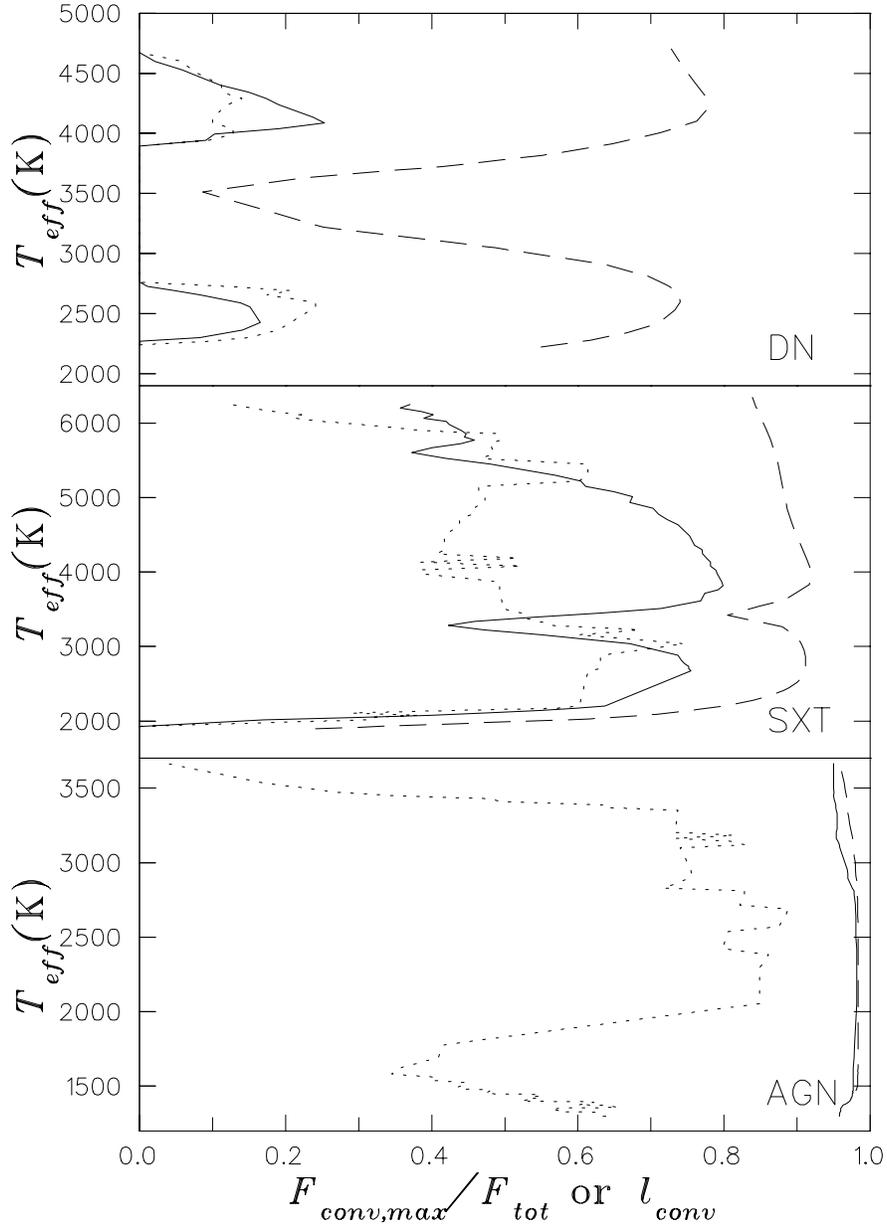}
\caption{Maximum values of convective fluxes
and vertical ranges of damped convection are plotted with
respect to effective temperature $T_{eff}$ for different
systems. The dashed lines denote $F_{conv,max}/F_{all,nodamp}$,
where $F_{conv,max}$ is the maximum MLT convective flux at
a given $T_{eff}$. The solid lines denote
$F_{conv,max}/F_{all,damp}$,
where $F_{conv,max}$ is the maximum damped convective flux at
a given $T_{eff}$. The vertical ranges (scaled with $h$)
of damped convection are indicated by the dotted lines.
\label{conv.eps}}
\end{figure}

\begin{figure}
\epsscale{.8}
\plotone{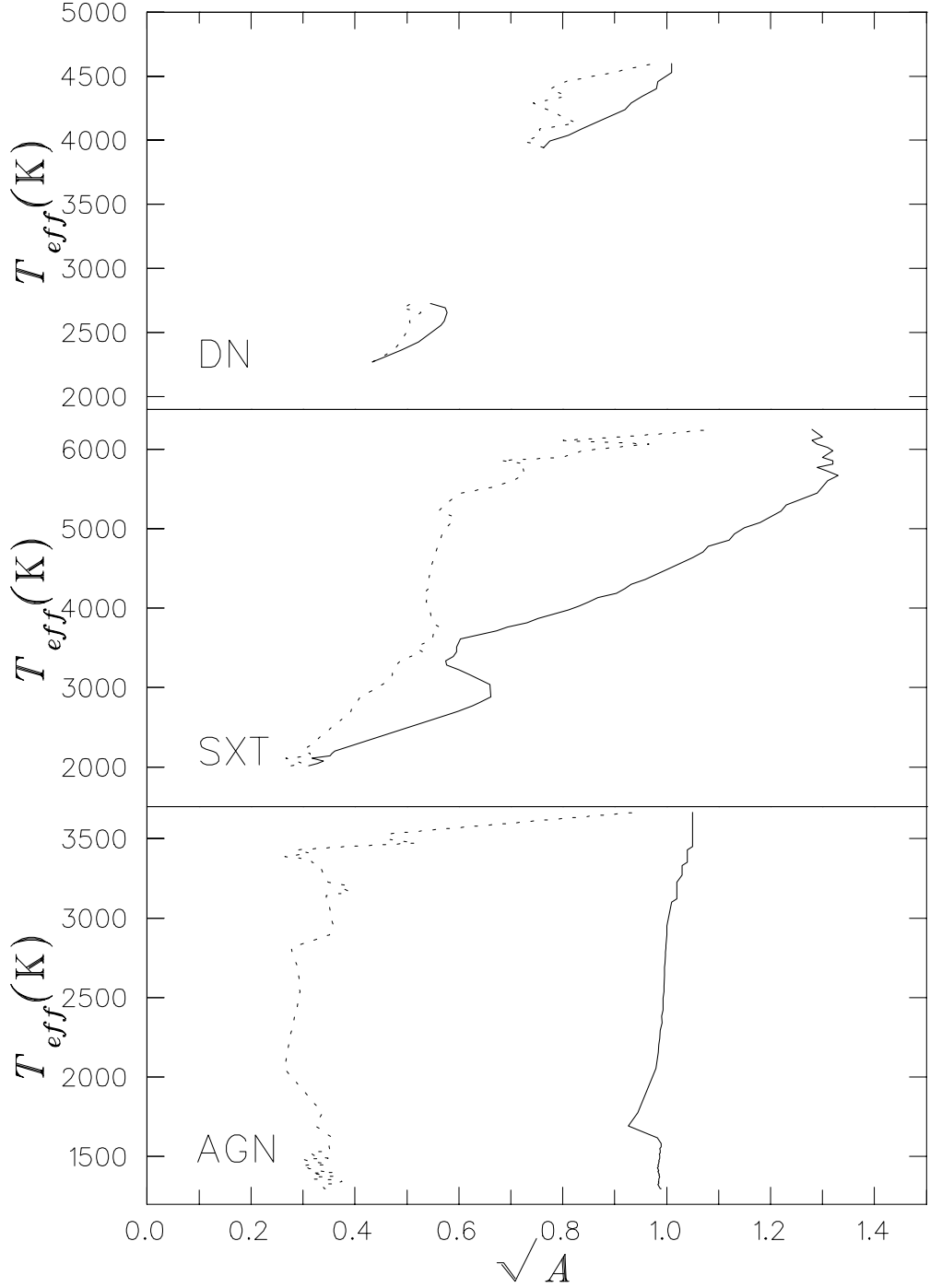}
\caption{Effective temperature vs. the ratio of radial
to vertical length $\sqrt{A}$ 
of damped convection for different
systems. The solid and dotted lines represent the maximal and
minimal values of $\sqrt{A}$ respectively at a given $T_{eff}$.
No solutions exist on the left side of dotted lines for weak
convection (i.e. smaller $\sqrt{A}$) as a result of strong damping.
\label{anisotropy.eps}}
\end{figure}

Fig.~\ref{conv.eps} and Fig.~\ref{anisotropy.eps}
illustrate the overall behavior of convective
flux along the unstable branches of different accretion disk systems. 
The ranges of effective temperature displayed in these figures are associated
with those for the thick solid lines (the curves $F_{all,damp}$) in
Fig.~\ref{fig1}, Fig.~\ref{fig3}, and Fig.~\ref{fig5}. Maximum values
of fractional convective fluxes (both MLT and damped) and the
vertical extent of the convective zones, $l_{conv}$ 
(scaled with half-thickness $h$), are depicted for damped convection
as a function of $T_{eff}$ in Fig.~\ref{conv.eps}.
MLT convection (dashed lines) becomes stronger and stronger
as one goes from DNe (top panel), to SXTs (middle panel), to 
AGN (bottom panel) and the difference between damped and MLT
convection diminishes over the same range.
The importance of damped convective flux can also be measured
by $l_{conv}$ (dotted lines), where $l_{conv}$ is the vertical 
range for which solutions to the minimization
problem exist (for which convection can actually occur). 
On average, $l_{conv}$ increases from $\approx 0.1$ for DNe to $\approx 0.8$
for AGN.  Fig.~\ref{anisotropy.eps} displays the maximal and minimal
values of $\sqrt{A}$ at a given $T_{eff}$ associated with the
S-curves for $F_{all,damp}$ (denoted by thick solid lines in
figures~\ref{fig1}, ~\ref{fig3}, and ~\ref{fig5})
for different systems. As mentioned above, $A$ tends to increase
with height and so a larger separation between maximal
and minimal $A$ goes with a larger vertical range for
damped convection. Values of $\sqrt{A}$ as small as $\sim 0.3$ can exist
for AGN but not for DNe in accord with the result that
convection is strong in AGN. In fact, Table~\ref{tb-DN_conv}
and Table~\ref{tb-AGN_conv} show that the convective growth
rate $\gamma \approx 0.1$ for DN at $\sqrt{A} \approx 0.8$
is comparable to that for AGN at $\sqrt{A} \approx 0.3$.
No solutions exist on the left side of dotted lines for weak
convection as a result of the severe damping of narrower convective
cells. On the other hand, `fatter' ($\sqrt{A}>1$)
convective eddies can exist for strong convection, since these
solutions reduce radiative losses and radial mixing and
extract only a small penalty in the form of increased
vulnerability to secondary instabilities.

\begin{figure}
\epsscale{.7}
\plotone{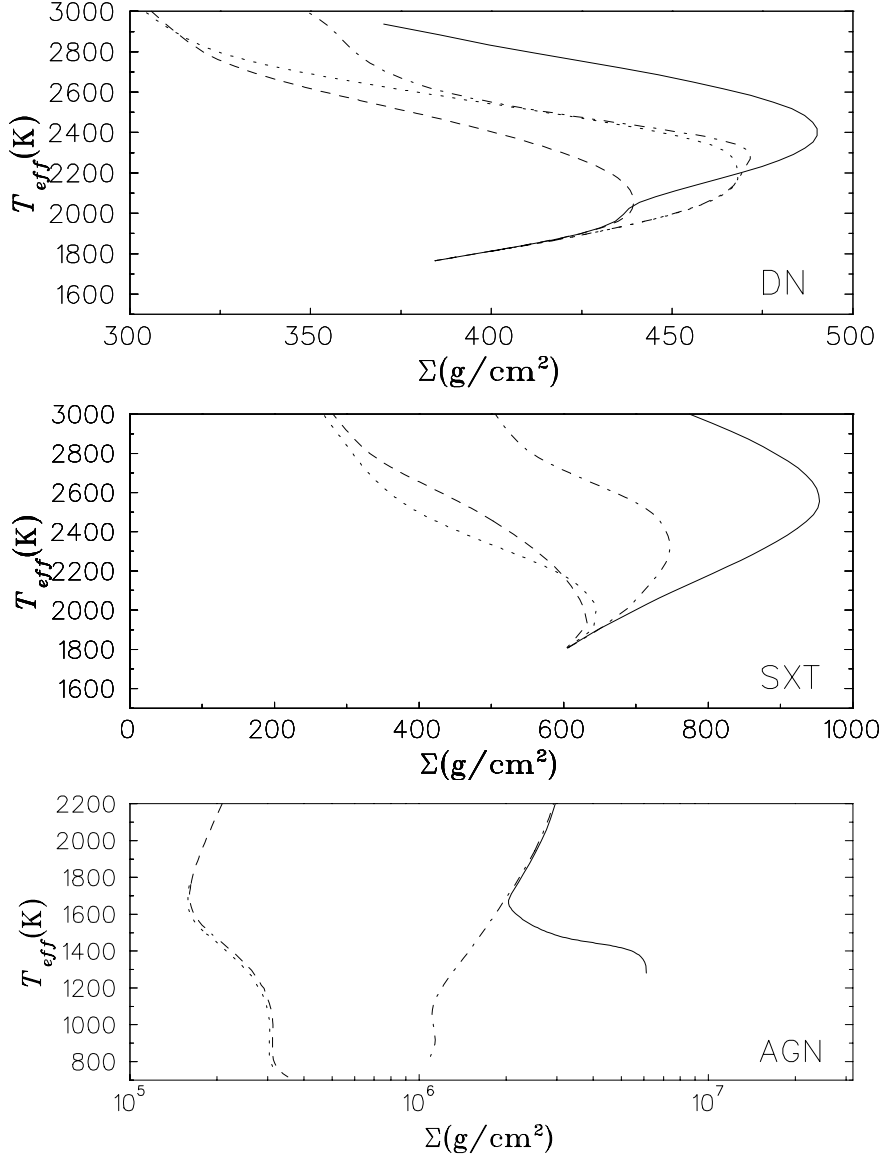}
\caption{Influence of metal abundance on the
S-curves around $\Sigma{max}$ for MLT convection and turbulent
mixing in DN (top panel), SXT (middle panel), and AGN
(bottom panel) disks. Solid lines indicate the curves
$F_{tot}=F_{rad}+F_{conv,MLT}$ with metals.
Dashed-dotted lines indicate the curves
$F_{tot}=F_{rad}+F_{conv,MLT}$ without metals.
Dashed lines denote the curves
$F_{tot}=F_{rad}+F_{turb}$ with metals.
And dotted lines denote the curves
$F_{tot}=F_{rad}+F_{turb}$ without metals.
\label{metal.fig}}
\end{figure}

Metals are the main supply of free electrons in the cold, neutral
accretion disk state.  In our code the ratio of electron gas pressure to the
total gas $P_e/P_g$ is $\sim 10^{-5}$ around $\Sigma_{max}$, indicating
that the disks are only weakly ionized. We calculated the 
microscopic resistivities due to the electron-neutral collisions
using the equation $\eta=230(n_n/n_e)T^{1/2}\,{\rm cm^2}\,{\rm s^{-1}}$
(\cite{bb94}) and used the standard formula for electron-ion collisions,
$\eta=2.65\times 10^{12} T^{-3/2} cm^2/s$. We found
magnetic Reynolds numbers 
\footnote{The more formal definition of $Re_M$ for
the isotropic turbulence driven by BH instability is
$\nu /\eta$.}
$Re_M\equiv \lambda_p c_s/\eta$
in quiescent DN and SXT disks are less than $10^4$, in accord
with previous work (\cite{gm98}), although this neglects the
possibility of ionizing particles from the disk coronae.
Fig.~\ref{metal.fig} illustrates how the S-curves
for disks
with solar metal abundances differ from ones without any metals around
$\Sigma_{max}$ for different systems. In order to distinguish the
effects of MLT convection from turbulent mixing, we use the curves
for ($F_{rad}+F_{conv,MLT}$) and ($F_{rad}+F_{turb}$).
For curves with MLT convection, the discrepancy between metal and no metal
models near cases around $\Sigma_{max}$
increases as one goes from DNe, to SXTs, to AGNs. This
is because convective mixing becomes stronger in AGNs. On the other hand,
the opposite trend happens to the curves with turbulent mixing
which follows from the fact that
turbulent mixing becomes weaker in AGN.

\begin{figure}
\epsscale{.8}
\plotone{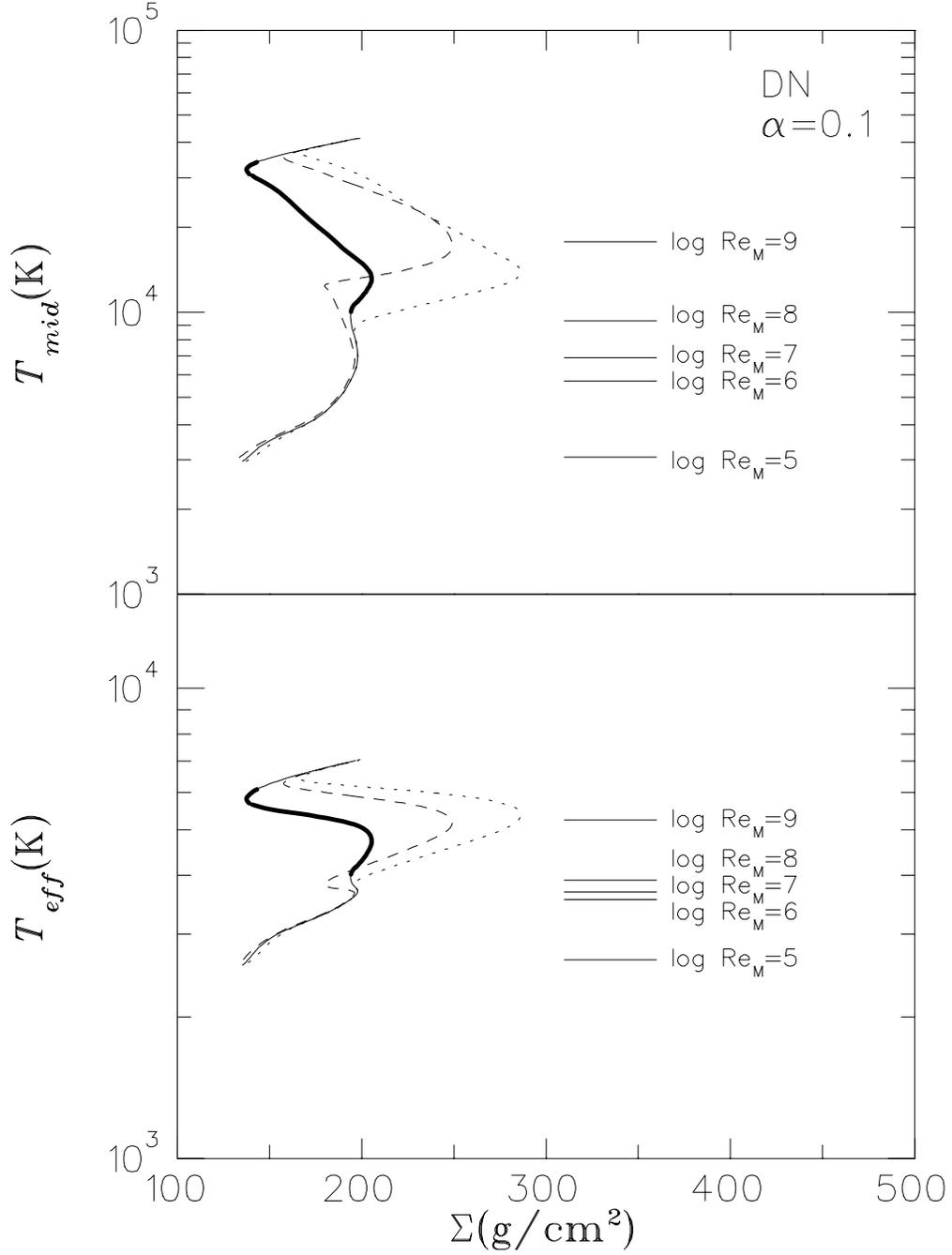}
\caption{S-curves of DN disks ($M_{WD}=\msun$,
$\log\,r({\rm cm})=9$) associated with different energy transports
for $\alpha=0.1$ around $\Sigma_{min}$:
$F_{all,damp}$ (thicker solid line), $F_{all,nodamp}$
(dotted line), $F_{rad}+F_{conv,MLT}$ (dashed line), and
$F_{rad}+F_{turb}$ (solid line).}
\label{s_curve_DNconst.eps}
\end{figure}

Finally, in Fig.~ (\ref{s_curve_DNconst.eps}) we test the consequences
of using a constant value of $\alpha$ instead of the scaling law.
This is motivated by Gammie and Menou's suggestion (\cite{gm98},
see also \cite{alp96})
that the timing of DN outbursts can be explained by the relatively
high resistivity of quiescent DN accretion disks.  If the BH
instability shuts off below $Re_M\sim 10^4$, then the low values of
$\alpha$ inferred from the intervals between outbursts reflect simply
the operation of some different angular momentum transport mechanism
in quiescent disks, and not a general scaling law.  
Fig.~ (\ref{s_curve_DNconst.eps}) shows the S-curve down to relatively
low temperatures, and the associated range of $Re_M$.  The
important feature here is that a cutoff of $Re_M\sim 10^5$ or less
results in 
S-curves with a very small range in column density.  While $\Sigma_{max}$
may still be very large if $\alpha$ drops sharply at still smaller
temperatures, the propagation of a cooling front
in an accretion disk depends to a large extent on the nature of the
S-curve near $\Sigma_{min}$ and detailed models of cooling fronts
indicate that the cooling sections of the disk trace a path just
below the unstable equilibrium curve (\cite{mhs99}).  A cooling
wave with a nearly
vertical drop, as seen here, cannot be described by the
rapid cooling wave theory (\cite{vw96},\cite{vis97}) but should
instead propagate at a fraction of the viscous accretion speed, just
as expected for an S-curve with a small value of 
$\Sigma_{max}/\Sigma_{min}$ (\cite{v98}).  In fact, models with constant $\alpha$
not only show slow cooling front speeds but also frequent
reversals (\cite{s84}, \cite{mhs99}).  Such models lack
a sharp drop in $\alpha$ at very low temperatures and may not
represent realistic calculations of the results of Gammie and
Menou's model.  However, it is not clear that even the existence of 
a second unstable branch at very low temperatures would affect the
cycling of a DN disk. The S-curve shown here is sufficient
to produce a complete, albeit weak, thermal cycle without ever 
moving to lower temperatures.  We caution however that
this conclusions depends on using a particular value of $Re_M$
for a cutoff. If the cutoff value of $Re_M$ lies in the
range $10^6$ to $10^7$ then a robust thermal limit cycle is 
still be possible for a constant $\alpha$ model. 

\section{Summary and Conclusions}

We have proposed a new model for vertical energy transport
in optically thick, Keplerian accretion disks which
takes into account the effects of turbulent mixing due to 
the BH instability.  This includes both an evaluation of
the turbulent flux $F_{turb}$ driven by approximately 
isotropic turbulent cells with an effective vertical thermal 
conductivity $\approx \nu$, and the increased damping of
the sheared convective cells.  The latter effect includes not only
radiative losses, which have been broadly discussed in the
literature, but also turbulent mixing which enhances heat
transport and retards convective motion.
Simple physical arguments, and the linear dispersion relation,
indicate that weak ($\Gamma_c<\Omega$) convective eddies
are radially thin in the shearing flow of an accretion disks. 
This enhances the effects of turbulent drag,
radial mixing, and radiative losses.  We account for these
effects through an extension of linear perturbation
theory using a one mode model.  We choose the linear mode
most resistant to secondary instabilities
and to turbulent damping and by calibrate the effects of this mode 
by requiring agreement with MLT if turbulent viscosity and shearing are 
removed. Following this procedure, our model for convection
becomes a minimization problem for the 
function $P_{tot}$ subject to the linear dispersion relation
involving several damping mechanisms.

Our use of quasilinear theory, calculating quadratic transport 
quantities and typical eddy shapes using linear modes, is
reasonable, but not rigorously justifiable.  Generally this
procedure works only when energy flows through a system in
ways that already appear in a linear analysis.  The appearance
of strong finite amplitude instabilities would destroy the
rationale behind this approach.  Fortunately thin accretion
disks do not appear to possess such instabilities (\cite{HBW99}).
One of the other major components of our approach, the suppression
of one family of unstable modes (convection, in this case) by
another is unusual, but not unprecedented.  In laboratory
plasmas we have the example of radial convection being
disrupted by the spontaneous appearance of shear (cf.
Burrell 1997).  In the context of accretion disks the suppression
of the Parker instability by BH induced turbulence
(\cite{SHGB96}), and previously predicted in Vishniac
and Diamond (1992), is very closely analogous to suppression
of convection.  The most problematic aspect of our treatment
is our method for choosing a `typical' linear mode (cf. 
equation (\ref{criterion})).  As far as we know this method has
never been proposed before and there are no definitive tests
of its usefulness.  However, it does capture the
correct limits, i.e. settling on the remaining unstable
mode in the case of marginal stability and choosing the
modes least susceptible to secondary Kelvin-Helmholtz
instabilities when damping is unimportant.  We conclude that
it is at least qualitatively correct.  

The thermal equilibrium S-curves calculated in our model differ
from previous models in several ways.  Our models tend to
reproduce previous work along the hot stable branch of the S-curve,
and give the same, or nearly the same $\Sigma_{min}$.  However,
the unstable branch tends to lie somewhat below, and to the left,
of previous work in the $\Sigma-T$ plane.  In addition, the
short, intermediate stable branch seen in previous work often
disappears, and is usually more nearly vertical when it does appear.
The value of $\Sigma_{max}$ is smaller in our models, though only
by 20-30\%.  Crudely speaking, these effects are stronger
for thicker disks, that is, for DNe and SXTs, rather than AGN.
However, in some respects these effects are more dramatic for
SXTs than for DNe, largely because the different transport
mechanisms give rise to more divergent S-curves in the former.
AGN disks are still reasonably well described by standard
MLT theory without turbulent mixing. 
There are two reasons for the way turbulent effects scale from
DNe to AGN.  First, in an average sense, convection becomes
stronger as we go to thinner disks around more massive objects.
$F_{conv}/F_{rad} \propto \Omega \Sigma^2$.  This leads to
faster convective growth rates, which means that convective
cells become more resistant to disruption by any reasonable level of
local turbulence.  Second, we have adopted a scaling law which
reinforces this trend by lowering the value of $\alpha$, and
consequently the strength of turbulent dissipation, in
thinner disks.  Since $\alpha$ decreases
and $\Sigma$ increases as one progresses from
DNe, to SXTs, to AGNs, we find an almost complete
suppression of convection in DNe. Models with no convection
are reasonably good approximations to our models with damped
convection. 

The specific calculations presented here rely to large extent
on rather uncertain approximations in extending MLT to strongly
sheared environments and in estimating the level turbulent
dissipation in disks.  While the procedure followed here is
reasonable enough, it would clearly be better to have means of
calibrating our theory for anisotropic convection.  Nevertheless,
the qualitative nature of our results are not sensitive to
the specific choices we have made.  On the whole, the results
are less sensitive to the exact level at which weak convection
is suppressed than to the qualitative point that weak convection
is vulnerable to turbulent mixing in a sheared environment.

The work presented here is based on the assumption that 
the scaling law equation (\ref{eq:ascale}) for $\alpha$ is
a reasonable approximation.  This tends to make our corrections
to MLT less important for thin disks, especially AGN disks.
In fact, the status of this scaling law is somewhat uncertain,
for reasons given above.  However, we note that a constant
$\alpha$, down to some limiting value of the magnetic Reynolds
number, seems unable to reproduce the basic phenomenology of
DNe outbursts.  At the same time a very low $\alpha$ seems
required from observations of quiescent DNe (for example 
\cite{woo86}).  
Unless the cooling wave theory is discarded altogther
it seems necessary to postulate some sort of geometric dependence
for $\alpha$.  On the other hand, the Bailey relation for DNe suggests that
the scaling law saturates for values of $\alpha$ typical of DNe
in outburst.  Finally, we note that 
the role of irradiation (\cite{kr98}) and possible
advection flows (\cite{nmy96}) in SXTs remain unclear, and prevent
a clear understanding of the form of $\alpha$ required to
explain observations of thin disks.

\acknowledgements

We thank Jaw-Luen Tang and Wei-Jr Wu for their assistance
in the code. We thank the anonymous referee for valuable comments.
We are grateful to Craig Wheeler and Stefano
Migliuolo for informative discussions. 
We also appreciate the prompt help of
Paola D'Alessio in obtaining his PhD thesis on the internet.
Part of this work was completed when one of us
(PG) was a visitor at the High Energy Physics Laboratory
in University of Texas at Austin, and he would like to thank
Roy Schwitters for his generous hospitality.  This work
was supported in part by NASA grant NAG5-2773.

\clearpage
\begin{deluxetable}{crrrrrrrr}
\footnotesize
\tablecaption{The Structure of Damped Convection
in DN disks: $\log\,h\,({\rm cm})=8.80$, $\log\,r\,({\rm cm})=10.5$,
$\alpha=8.89\times 10^{-2}$.\label{tb-DN_conv}}
\tablewidth{0pt}
\tablehead{
\colhead{$z$} & \colhead{$\Gamma'$} & \colhead{$\gamma/\Omega$} &
\colhead{$\sqrt{A}$} & \colhead{$P_{tot}$} &
\colhead{$\nabla/\nabla'$} &
\colhead{$v_z/c_s$} & \colhead{$k^2\nu /\Omega$} &
\colhead{$1/(1+a_0\Gamma'')$} 
}
\startdata
0.67 & 2.01(5.01) & 0.07(0.51) & 0.81 & 117 & 1.10(2.07)
     & 0.03(0.21) & 0.76 & 0.06 \nl
0.70 & 1.72(3.29) & 0.12(0.56) & 0.88 &  45 & 1.16(1.98) 
     & 0.05(0.22) & 0.77 & 0.11 \nl
0.72 & 1.02(1.57) & 0.16(0.61) & 0.95 &  26 & 1.18(1.70) 
     & 0.06(0.23) & 0.81 & 0.21 \nl
0.74 & 0.12(1.57) & 0.06(0.58) & 0.91 & 249 & 1.04(1.44) 
     & 0.02(0.21) & 0.99 & 0.42 \nl
\enddata
\end{deluxetable}

\begin{deluxetable}{crr}
\footnotesize
\tablecaption{Values of $\alpha$ in the upper, middle, and
lower branches of S-curves for DN disks
in Fig.~\ref{fig1}.\label{tb-DN_alpha}}
\tablewidth{0pt}
\tablehead{
\colhead{$\alpha_0$} & \colhead{branch} & \colhead{$\alpha$}
}
\startdata
50 & upper & $\approx 0.25-0.61$ \nl
   & middle& $\approx 0.02-0.25$ \nl
   & lower & $\lesssim 0.02$     \nl
20 & upper & $\approx 0.11-0.26$ \nl
   & middle& $\approx 0.007-0.11$\nl
   & lower & $\lesssim 0.007$    \nl
\enddata
\end{deluxetable}

\begin{deluxetable}{crrrrrrrr}
\footnotesize
\tablecaption{The Structure of Damped Convection
in LMXB disks: $\log\,h\,{\rm cm}=8.33$, $\log\,r\,{\rm cm}=10.5$,
$\alpha=1.55\times 10^{-2}$.\label{tb-LMXB_conv}}
\tablewidth{0pt}
\tablehead{
\colhead{$z$} & \colhead{$\Gamma'$} & \colhead{$\gamma/\Omega$} &
\colhead{$\sqrt{A}$} & \colhead{$P_{tot}$} &
\colhead{$\nabla/\nabla'$} &
\colhead{$v_z/c_s$} & \colhead{$k^2\nu /\Omega$} &
\colhead{$1/(1+a_0\Gamma'')$} 
}
\startdata
0.47 & 34.5(83.5) & 0.07(0.21) & 0.55 &  23 & 1.25(1.51) 
        & 0.03(0.09) & 0.30 & 0.01 \nl
0.66 & 14.4(15.7) & 0.18(0.36) & 0.64 & 3.2 & 1.49(1.50) 
        & 0.07(0.14) & 0.21 & 0.09 \nl
0.72 & 6.63(6.95) & 0.25(0.44) & 0.72 & 2.4 & 1.53(1.52) 
        & 0.09(0.15) & 0.22 & 0.20 \nl
0.79 & 1.11(1.35) & 0.33(0.52) & 0.81 & 2.0 & 1.35(1.36) 
        & 0.10(0.15) & 0.28 & 0.61 \nl
\enddata
\end{deluxetable}

\begin{deluxetable}{crrr}
\footnotesize
\tablecaption{Values of $\alpha$ in the upper, middle, and
lower branches of S-curves for SXT disks
in Fig.~\ref{fig3} and Fig.~\ref{fig3.9}.\label{tb-LMXB9_alpha}}
\tablewidth{0pt}
\tablehead{
\colhead{$\alpha_0$} & \colhead{branch} & \colhead{$\alpha$ at
$\log\,r\,({\rm cm})=10.5$} & \colhead{$\alpha$ at
$\log\,r\,({\rm cm})=9$}
}
\startdata
50 & upper & $\approx 0.06-0.15$ & $\gtrsim 0.004$ \nl
   & middle& $\approx 0.003-0.06$ & $\approx 0.0002-0.004$ \nl
   & lower & $\lesssim 0.003$     & $\lesssim 0.0002$  \nl
\enddata
\end{deluxetable}

\begin{deluxetable}{crrrrrrrr}
\footnotesize
\tablecaption{The Structure of Damped Convection
in AGN disks: $\log\,h\,({\rm cm})=12.58$, $\log\,r\,({\rm cm})=15.5$,
$\alpha=8.32\times 10^{-4}$.\label{tb-AGN_conv}}
\tablewidth{0pt}
\tablehead{
\colhead{$z$} & \colhead{$\Gamma'$} & \colhead{$\gamma/\Omega$} &
\colhead{$\sqrt{A}$} & \colhead{$P_{tot}$} &
\colhead{$\nabla/\nabla'$} &
\colhead{$v_z/c_s$} & \colhead{$k^2\nu /\Omega$} &
\colhead{$1/(1+a_0\Gamma'')$} 
}
\startdata
0.28 & 7.73$\times 10^5$(7.40$\times 10^6$) & 0.05(0.11) 
        & 0.28 & 10.8 & 1.06(1.03) & 0.01(0.03) 
        & 9.92$\times 10^{-2}$ & 1.70$\times 10^{-6}$ \nl
0.46 & 4.91$\times 10^5$(2.54$\times 10^6$) & 0.11(0.16) 
        & 0.35 & 4.6 & 1.07(1.05) & 0.03(0.04) 
        & 9.45$\times 10^{-2}$ & 5.61$\times 10^{-6}$ \nl
0.66 & 9.33$\times 10^4$(2.46$\times 10^5$) & 0.28(0.41) 
        & 0.56 & 2.1 & 1.15(1.12) & 0.05(0.06) 
        & 8.81$\times 10^{-2}$ & 5.94$\times 10^{-5}$ \nl
0.85 & 1.66$\times 10^3$(3.05$\times 10^3$) & 0.93(1.12)
        & 0.90 & 1.13 & 1.60(1.56) & 0.11(0.12) 
        & 9.66$\times 10^{-2}$ & 7.90$\times 10^{-3}$ \nl
0.93 & 1.63(2.84) & 2.99(3.52) 
        & 0.99 & 1.02 & 1.80(2.25) & 0.24(0.27)
        & 1.68$\times 10^{-1}$ & 9.36$\times 10^{-1}$ \nl
\enddata
\end{deluxetable}

\begin{deluxetable}{crr}
\footnotesize
\tablecaption{Values of $\alpha$ in the upper, middle, and
lower branches of S-curves for AGN disks
in Fig.~\ref{fig5}.\label{tb-AGN_alpha}}
\tablewidth{0pt}
\tablehead{
\colhead{$\alpha_0$} & \colhead{$T_{mid}$} & \colhead{$\alpha$}
}
\startdata
50 & $80000-210000$ & $\approx 0.004-0.02$ \nl
   & $37000-80000$ & $\approx 0.0018-0.004$ \nl
   & $14000-37000$ & $\approx 0.0006-0.0018$ \nl
   & $5000-14000$  & $\approx 0.0002-0.0006$ \nl
\enddata
\end{deluxetable}

\clearpage

\clearpage

\begin{thebibliography}{}
\bibitem[Alexander, Johnson, and Rypma 1983]{ajr83}
    Alexander, D. R., Johnson, H. R., and Rypma, R. L.
    1983, \apj, 272, 773
\bibitem[Armitage, Livio, and Pringle  1996]{alp96}
    Armitage, P. J., Livio, M., and Pringle, J. E. 1996,
    \apj, 457, 332
\bibitem[Balbus and Hawley 1991]{bh91}Balbus, S.,
and Hawley, J.F.\ 1991, \apj, 376, 214
\bibitem[Balbus and Hawley 1998]{bh98} Balbus, S.,
    and Hawley, J. F. 1998, Rev. Mod. Phys., 70, 1
\bibitem[Batchelor 1950]{b50} Batchelor, G.K.\ 1950,
Proc. Roy. Phil. Lond., A201, 405
\bibitem[Blaes and Balbus 1994]{bb94} Blaes, O. M.,
    and Balbus, S. A. 1994, \apj, 421, 163
\bibitem[Burrell\ 1997]{Bu97}Burrell, K.H.\ 1997, Phys. Plasmas, 4, 1499
\bibitem[Cabot et. al. 1987]{cchp87} Cabot, W.,
    Canuto, V. M., Hubickyj, O., \& Pollack, J. B. 1987,
    Icarus, 69, 423
\bibitem[Cannizzo 1992]{can92} Cannizzo, J. K. 1992,
    \apj, 385, 94
\bibitem[Cannizzo 1993]{can93} Cannizzo, J. K. 1993,
    in Accretion Disks in Compact Stellar Systems, ed.
    J. C. Wheeler (Singapore: World Scientific), p. 6
\bibitem[Cannizzo 1994]{can94} Cannizzo, J. K. 1994,
    \apj, 435, 389
\bibitem[Cannizzo and Cameron 1988]{cc88} Cannizzo, J. K.,
    and Cameron, A. G. W.  1988, \apj, 330, 327
\bibitem[Cannizzo, Chen, and Livio 1995]{ccl95} Cannizzo, J. K.,
    Chen, W., and Livio, M. 1995, \apj, 454, 880
\bibitem[Cannizzo and Wheeler 1984]{cw84} Cannizzo, J. K.,
    and Wheeler, J. C. 1984, \apjs, 55, 367
\bibitem[Canuto and Hartke 1986]{ch86} Canuto V. M.,
    and Hartke, G. J. 1986, \aap, 168, 89
\bibitem[Chandrasekhar, 1960]{chan60} Chandrasekhar, S. 1960,
    Proc, Nat. Acad. Sci., 46, 253
\bibitem[Chen, Shrader, and Livio 1997]{csl97} Chen, W.,
    Shrader, C. R., and Livio, M. 1997, \apj, 491, 312
\bibitem[Cox and Giuli 1968]{cox68} Cox, J. P.,
    and Giuli, R. T. 1968, Principles of Stellar
    Structure (New York: Gordon and Breach)
\bibitem[Cox and Stewart 1969]{cs69} Cox, A. N.,
    and Stewart, J. N. 1969,
    Sci. Inf., Astr. Council, Acad. Sci. USSR, No. 15, p. 1.
\bibitem[D'Alessio et. al. 1998]{DAlessio} D'Alessio, P.,
    Cant\'o, J., Calvet, N., and Lizano, S. 1998, \apj, 500, 411
\bibitem[Dubus et al. 1999]{dlhc99}Dubus, G., La-Sota, J.P., Hameury,
    J.-M., and Charles, P.A.\ 1999, \mnras, 302, 731
\bibitem[Gammie\ 1997]{g97}Gammie, C.F.\ 1997, astro-ph/9712233
\bibitem[Gammie and Menou 1998]{gm98} Gammie, C.F.,
    and Menou, K. 1998, \apj, 492, L75
\bibitem[Hameury et al.\ 1998]{hmd98}Hameury, J.-M., Menou, K.,
Dubus, G., Lasota, J.-P., and Hure, J.-M.\ 1998, \mnras, 298, 1048
\bibitem[Hawley and Balbus 1992]{hb92} Hawley, J.,
    and Balbus, S. 1992, \apj, 400, 595
\bibitem[Hawley, Balbus and Winters\ 1999]{HBW99}Hawley, J.F.,
Balbus, S.A., and Winters, W.F.\ 1999, \apj, 518, 394
\bibitem[Honma\ 1996]{h96} Honma, F.\ 1996, \pasj, 48, 77
\bibitem[Iben 1963]{ibe63} Iben, Jr., I. 1995,
    \apj, 138, 452
\bibitem[Kazantsev\ 1967]{k67}Kazantsev, A.P.\ 1967, JETP, 53, 1806
\bibitem[King \& Ritter\ 1998]{kr98}King, A.R., and Ritter, H.\ 1998,
    \mnras, 293, 42
\bibitem[Lin and Shields 1986]{ls86} Lin, D. N. C., and
    Shields, G. A. 1986, \apj, 305, 28
\bibitem[Livio and Spruit 1991]{ls91} Livio, M., and
    Spruit, H. C. 1991, \aap, 252, 189
\bibitem[Ludwig, Meyer-Hofmeister, and Ritter 1994]{lmr94}
    Ludwig, K., Meyer-Hofmeister, E., and Ritter, H. 1994,
    \aap, 290, 473
\bibitem[Maeder 1995]{mae95} Maeder, A. 1995,
    \aap, 299, 84
\bibitem[Maeder and Meynet 1996]{mae96} Maeder, A.,
    and Meynet G. 1995, \aap, 313, 140
\bibitem[Menou, Hameury, \& Stehle\ 1999]{mhs99}
Menou, K., Hameury, J.-M., and Stehle, R.\ 1999, \mnras,
in press
\bibitem[Meyer and Meyer-Hofmeister 1982]{mm82} Meyer, F.,
    and Meyer-Hofmeister E. 1982, \aap, 106, 34
\bibitem[Mineshige 1988]{min88} Mineshige, S. 1988, \aap, 190, 72
\bibitem[Mineshige and Osaki 1983]{mo83} Mineshige, S.,
    and Osaki, Y. 1983, \pasj, 35, 377
\bibitem[Narayan, McClintock, and Yi 1996]{nmy96} Narayan, R.,
    McClintock, J. E., and Yi, I. 1996, \apj, 457, 821
\bibitem[Osaki 1996]{osa96} Osaki, Y. 1996, \pasp, 108, 39
\bibitem[Pojma\'nski 1986]{poj86} Pojma\'nski, G. 1986,
    Acta Astr., 36, 69
\bibitem[Pollack, McKay and Christofferson 1985]{pmc85}
    Pollack, J. B., McKay, C. P., and Christofferson, B. M.
    1985, Icarus, 64, 471
\bibitem[Ruden, Papaloizou, and Lin 1988]{rpl88} Ruden, S. P.,
    Papaloizou, J. C. B., \& Lin, D. N. C. 1988, \apj, 329, 739
\bibitem[R\"udiger et. al. 1988]{rud} R\"udiger, C., Elstner, D.,
    and Tsch\"ape, R., 1988, Acta Astr., 38, 299
\bibitem[Sakimoto and Coroniti 1981]{sc81} Sakimoto , P. J.,
    and Coroniti, F. V., 1981, \apj, 247, 19
\bibitem[Schittkowski 1985]{sch85} Schittkowski, K. 1985/86,
    Annals of Operations Research, 5, 4850
\bibitem[Shakura and Sunyaev 1973]{ss73} Shakura, N. I.,
    and Sunyaev, R. A. 1973, \aap, 24, 337
\bibitem[Smak 1984]{s84} Smak, J. 1984,
    Acta Astr., 34, 161
\bibitem[Spellucci 1998]{spe98} Spellucci, P. 1998,
    URL: http://plato.la.asu.edu/guide.html
\bibitem[Speziale, Sarkar and Gatski\ 1991]{ssg91} Speziale, C.G.,
Sarkar, S., and Gatski, T.B.\ 1991, J. Fluid Mech., 227, 245
\bibitem[Speziale and Gatski\ 1997]{sg97} Speziale, C.G.,
Sarkar, S., and Gatski, T.B.\ 1997, J. Fluid Mech., 344, 155
\bibitem[Stone et. al. 1996]{SHGB96}
Stone, J., Hawley, J.F., Gammie, C.F., and Balbus, S.A.\ 1996, \apj,
463, 656
\bibitem[Talon and Zahn\ 1997]{tz97} Talon, S., and Zahn, J.-P.
1997, \aap, 317, 749
\bibitem[Velikhov 1959]{vel59} Velikhov, E. P. 1959,
    Soviet Phys. - JETP, 36, 1398
\bibitem[Vishniac 1993]{vis93} Vishniac, E. T. 1993,
    in Accretion Disks in Compact Stellar Systems, ed.
    J. C. Wheeler (Singapore: World Scientific), p. 41
\bibitem[Vishniac 1997]{vis97} Vishniac, E. T. 1997,
    \apj, 482, 414
\bibitem[Vishniac 1998]{v98}Vishniac, E.T.\ 1998,
Proceedings of the 1997 Maryland Meeting, in press
(available as astro-ph/9802232)
\bibitem[Vishniac 1999]{v99} Vishniac, E.T.\ 1999,
proceedings of the 1998 Chapman Conference on Magnetic
Helicity, ed. A. Pevtsov (AGU: New York)
\bibitem[Vishniac and Diamond 1992]{vd92} Vishniac, E. T.,
    and Diamond, P. H. 1992, \apj, 398, 561
\bibitem[Vishniac and Brandenburg 1997]{vb97}Vishniac, E.T.,
    and Brandenburg, A.\ 1997, \apj, 475, 263
\bibitem[Vishniac and Wheeler 1996]{vw96} Vishniac, E. T.,
    and Wheeler, J. C. 1996, \apj, 471, 921
\bibitem[Wood et al. 1986, 1989]{woo86} Wood, J. H.,
    Horne, K., Berriman, G., \& Wade, R. A. 1989, \apj, 341, 974
\bibitem[Wood et al. 1986, 1989]{} Wood, J. H., Horne, K.,
    Berriman, G., Wade, R. A., \& O'Donoghue, D. 1986,
    \mnras, 219, 629
\bibitem[Zahn 1992]{zah92} Zahn, J.-P., 1992, \aap, 265, 115
\end{thebibliography}
\end{document}